\documentclass[lettersize,journal]{IEEEtran}
\IEEEoverridecommandlockouts
\usepackage[color=red]{todonotes}
\usepackage{tikz}
\usepackage{amsmath}

\usepackage{filecontents}

\usepackage{cite}
\usepackage{amsmath,amssymb,amsfonts}
\usepackage{algorithmic}
\usepackage{graphicx}
\usepackage{textcomp}
\usepackage{xcolor}
\usepackage{todonotes}
\PassOptionsToPackage{hyphens}{url}
\usepackage{hyperref}
\usepackage{multirow}
\usepackage{comment}
\usepackage{ulem}
\usepackage{subcaption}

\usepackage{array}

\def\BibTeX{{\rm B\kern-.05em{\sc i\kern-.025em b}\kern-.08em
    T\kern-.1667em\lower.7ex\hbox{E}\kern-.125emX}}

\begin{document}

\title{\Large \bf \textit {Impedance Leakage} Vulnerability and its Utilization in Reverse-engineering \\ Embedded Software}

%Uncovering Impedance Leakage: Reverse-Engineering Strategies for Embedded Software

%{Impedance Leakage} and its Exploitation in Reverse-engineering Embedded Software

\author{\IEEEauthorblockN{Md Sadik Awal and Md Tauhidur Rahman}\\
\IEEEauthorblockA{\textit{Security, Reliability, Low-power, and Privacy (SeRLoP) Research Lab \\ECE Department, Florida International University, Miami, Florida, USA} \\
%\textit{Florida International University, Miami, Florida, USA}
E-mail: \{mawal003 and mdtrahma\}@fiu.edu}
}

\begin{comment}
\author{Masahito Hayashi
\IEEEmembership{Fellow, IEEE}, Masaki Owari
\thanks{M. Hayashi is with Graduate School
of Mathematics, Nagoya University, Nagoya,
Japan}
\thanks{M. Owari is with the Faculty of
Informatics, Shizuoka University,
Hamamatsu, Shizuoka, Japan.}
}
\end{comment}

\maketitle
\begin{abstract}
% Protecting systems and data from physical attacks through unauthorized access is crucial. 
Discovering new vulnerabilities and implementing security and privacy measures are important to protect systems and data against physical attacks. % that exploit unauthorized access. Although impedance is commonly studied in antenna, radio-frequency, and biomedical research, its impact on privacy and security has been largely overlooked.
One such vulnerability is impedance, an inherent property of a device that can be exploited to leak information through an unintended side channel, thereby posing significant security and privacy risks. Unlike traditional vulnerabilities, impedance is often overlooked or narrowly explored, as it is typically treated as a fixed value at a specific frequency in research and design endeavors. 
Moreover, impedance has never been explored as a source of information leakage. 
This paper demonstrates that the impedance of an embedded device is not constant and directly relates to the programs executed on the device. We define this phenomenon as \textit{impedance leakage} and use this as a side channel to extract software instructions from protected memory. 
% Recent attacks on embedded devices emphasize the pressing need for a solution to protect against malware and maintain software privacy. Although there are several anomaly detection mechanisms, side channel signals have emerged as being very effective since they can monitor malicious activities or the secure execution of programs without disrupting the system under observation. Moreover, physical side channel-based monitoring provides a high level of security by being specialized for the architectures and programs they are applied to, reducing attack vectors, and monitoring immutable physical side channels using an undecipherable data-driven model. 
% Existing side channel-based instruction monitors examine the power traces and electromagnetic leakage of a device. However, they have a number of constraints, such as device modifications and extensive implementation requirements. 
Our experiment on the ATmega328P microcontroller and the Artix 7 FPGA indicates that the impedance side channel can detect software instructions with 96.1\% and 92.6\% accuracy, respectively. Furthermore, we explore the dual nature of the impedance side channel, highlighting the potential for beneficial purposes and the associated risk of intellectual property theft. Finally, potential countermeasures that specifically address impedance leakage are discussed.  

%In this research, a novel non-invasive physical side channel attack is introduced, leveraging the data-dependent changes in chip impedance. The attack capitalizes on the modifications to the circuit's physical characteristics caused by the contents temporarily stored in registers, resulting in variations in die impedance. To detect these impedance changes, scattering parameter analysis, a widely-used RF/microwave technique, is employed. By injecting high-frequency sine wave signals into the system's power distribution network (PDN) and measuring the signal echoes, the reflected signal is modulated differently at various frequency points based on the content bits and physical location of a register. This enables simultaneous and independent probing of individual registers. 
\end{abstract}


\section{Introduction} \noindent 
% modified
Integrated circuits and systems have become essential components in various domains, managing sensitive data and attracting adversaries aiming to extract unauthorized access or information \cite{yaacoub2020cyber}. 
Intriguingly, the internal operations of these systems give rise to physical side channel signals—subtle traces of information that can be exploited to obtain sensitive information. In this context, physical side channel analysis (SCA) arises as a potent technique based on the physical properties of a system \cite{standaert2010introduction}. The phenomenon of physical side channel leakages resulting from the computation and storage operations on integrated circuits reveals itself through measurable quantities such as power consumption, electromagnetic (EM) radiation, and even thermal radiation. The attackers exploit these quantities to compromise the security of cryptographic implementations \cite{standaert2010introduction}. Although numerous countermeasures have been proposed and developed to limit SCA-based attacks, SCA has demonstrated their effectiveness despite standard security measures \cite{brisfors2022side,zhou2005side,yao2018fault}. The persistence and adaptability of side channel vulnerabilities continue to pose new challenges \cite{panoff2022review,zhou2005side}.

Contrary to conventional security measures, which often focus on mitigating vulnerabilities, 
% there is a silver lining of SCA: it can also be employed to enhance hardware security. 
side channel analysis presents an intriguing perspective: It can be employed to enhance hardware security \cite{kong2009hardware,lin2009trojan,dunlap2016using,spatz2019review}. By integrating side channel information into device design, intrusion detectors can be developed to monitor the side channels of a device for indications of an attack. These detectors can identify deviations from expected patterns and initiate preventive measures if an attack is detected \cite{spatz2019review}. Moreover, external detectors can discreetly observe device activity without relying on its inputs and outputs, thus foiling an attacker's attempts to evade detection by controlling those elements. 

While many vulnerabilities resulting from the physical properties of devices are known, one often overlooked aspect is impedance—a fundamental property \cite{10133318}. Yet, this seemingly innocuous factor can be exploited as an unintended side channel, posing significant risks to security and privacy. Traditionally, impedance is treated as a fixed value at a specific frequency \cite{10133318,nguyen2019creating}, limiting its exploration. 
In contrast to the prevailing assumption, our research studies the relationship between device impedance and sensitive information. 
%However, challenging this assumption, our research delves into the relation between impedance and information leakage. 
We demonstrate that impedance serves as a side channel leaking sensitive information from an embedded device and define this phenomenon as \textit{impedance leakage} \cite{10133318,PAINE_Paper}. 
Subsequent research, such as that conducted by \cite{leakyohm}, follows our foundational contributions in recognizing impedance as a side channel to recover the AES encryption key. This provides a seamless continuity within the progression of research in this domain. 
In this study, conducted on two different hardware platforms, we use impedance leakage to extract information about the executed software instructions on the target devices. By shedding light on this overlooked characteristic, impedance, we open a new direction for understanding and securing systems in the face of emerging physical SCA.

In this article, we explore the potential of impedance leakage in reverse engineering software instructions. Disassembling software instructions provides insights into the internal behavior of an embedded device, making it a potential defense against a wide range of software-based attacks, e.g., software modification-based attacks. We discover that the impedance of an embedded device changes when it executes different software instructions. Leveraging this observation, we analyze and investigate the feasibility of utilizing device impedance as an emerging side channel for software instruction disassembly.
Thus, this work addresses the question: ``Does device impedance leak information of the executed software instructions and emerge as a promising side channel?”—or in other words: \textit{Can we use the device impedance to reverse engineering executed software instructions?} 
Our contributions in this paper are summarized as follows: 
\begin{itemize}
    \item  We explore a novel impedance side channel generated by the switching activities of a device. % \todo[inline]{We claimed this in the HOST paper. Should we modify or remove this?} 
    \item We present a unique approach to analyzing device impedance as a source of information leakage in extracting distinctive fingerprints for classifying program instructions. 
    \item We demonstrate the existence of impedance leakage by reverse engineering the software instructions on different embedded platforms.  
    % \item We find that the impedance side channel can identify different software instructions. 
\end{itemize}

The remaining sections of the paper are organized as follows. Section~\ref{sec:background} briefly addresses existing side channels along with impedance as a side channel. % and the vector network analyzer (VNA) in measuring the impedance signal. 
Section~\ref{sec:ILinReverseEngineering} discusses the research motivation. % The proposed framework steps are outlined in Section~\ref{sec:framework}. 
The experimental setup of this study has been discussed in Section~\ref{sec:experimental_setup}, which includes the hardware setup, software setup, and impedance signal measurement. In Section~\ref{sec:evaluation}, the analysis of the signals and the classification results are presented. Section~\ref{sec:related_work} addresses the related works. The conclusion of the work is provided in Section~\ref{sec:conclusion}.

\section{Background}
\label{sec:background} \noindent 
The security and privacy of digital systems are not solely dependent on the efficacy of mathematically secure algorithms and protocols, as the physical properties of a system can unintentionally leak internal information. These unintended physical characteristics, known as physical side channels, can be exploited by attackers to gain access to sensitive data and assets. Two popular types of physical side channels are power SCA and EM SCA. 
% Physical SCAs pose significant threats to the security of a system, as they inadvertently leak internal information. Among these, power SCA and EM SCA are prominent techniques used by attackers to exploit vulnerabilities.

Power SCA involves analyzing variations in power consumption during system operations. An attacker can deduce sensitive information like cryptographic keys by measuring the power consumption of a target component, such as a processor. Differential power analysis (DPA) is commonly used in this context to detect subtle changes in power consumption, providing insights into data-dependent variations that may otherwise be imperceptible. Alternatively, EM SCA relies on capturing the EM radiation emanating from a device during its functioning. As the device executes its operations, the dynamic changes in the current flow within its components give rise to distinct EM waves. Through the careful analysis of these EM leakages, attackers can glean valuable information about the events occurring throughout each clock cycle, allowing them to potentially infer internal processes and sensitive data. 

Despite their effectiveness, both power and EM SCAs have limitations. Power SCA requires the modification of the original device with a shunt resistor, which may not always be feasible or practical. Additionally, EM SCA can be challenging due to the need for specialized equipment like near-field probes or antennas and the necessary signal processing steps to extract exploitable information. To address these limitations and strengthen the security of digital systems against emerging physical SCAs, exploration is required to understand their unintended correlations with sensitive information.  
% Therefore, to enhance electronic system security, researchers explore new side channel avenues based on unexplored physical phenomena and correlations.

\subsection{Impedance Leakage}
\label{sec:imp_model} 
\begin{figure}[htbp]
\begin{subfigure}{.18\textwidth}
    \centering
    \includegraphics[width = 1\textwidth]{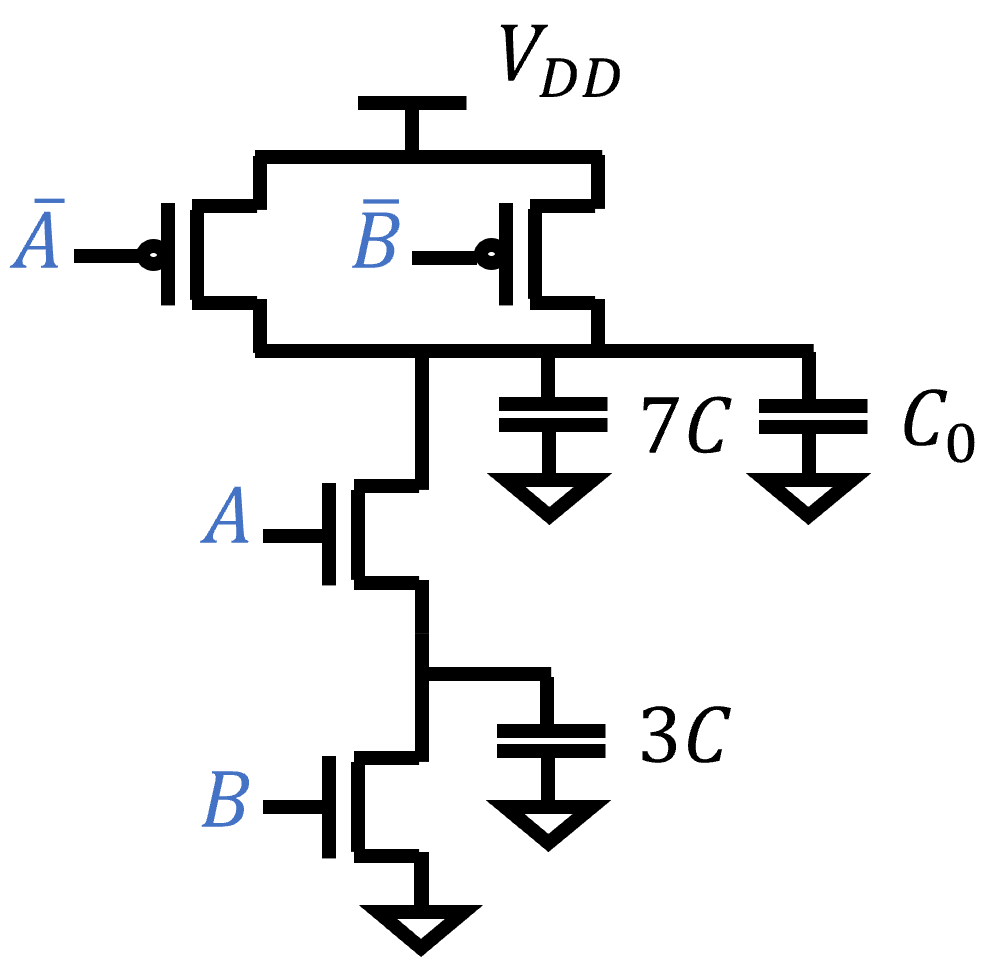}
    \caption{}
    \label{fig:nand_eq_single}
\end{subfigure}
\begin{subfigure}{.17\textwidth}
    \centering
    \includegraphics[width = 1\textwidth]{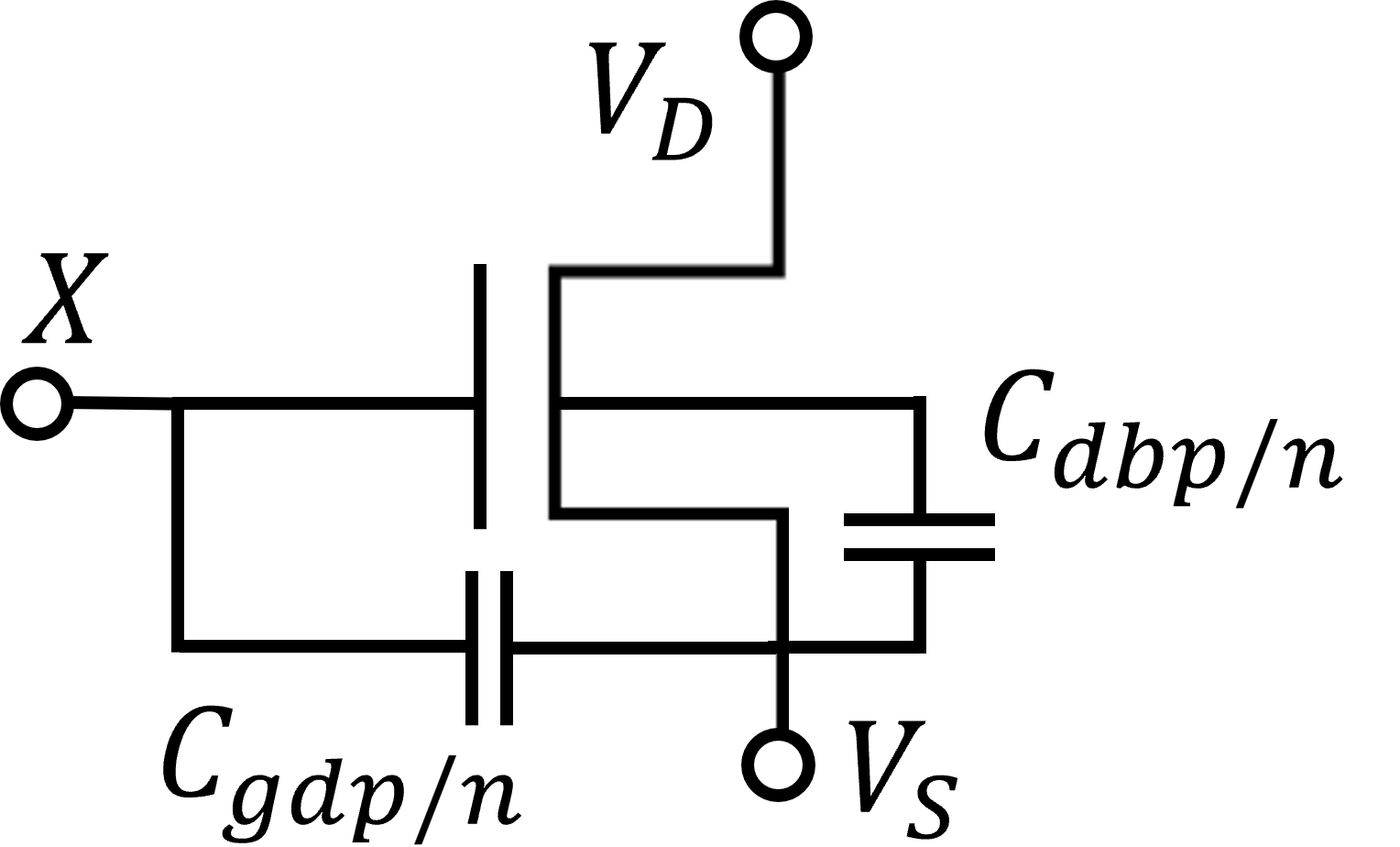}
    \caption{}
    \label{fig:mosfet_para_cap_single}
\end{subfigure}
\begin{subfigure}{.11\textwidth}
    \centering
    \includegraphics[width = 0.92\textwidth]{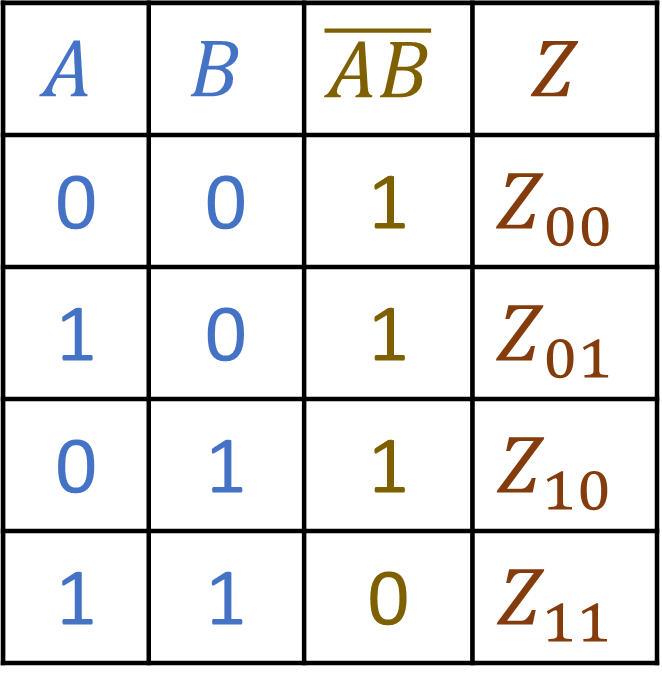}
    \caption{}
    \label{fig:truthtable_single}
\end{subfigure}
\caption{(a) 2-input NAND gate, (b) PMOS/NMOS with the parasitic capacitance, (c) truth table with relative impedance.}
\label{fig:nand_eq}
\end{figure}

\noindent 
\begin{comment}
\noindent
\textbf{Impedance Leakage at the transistor-level.}

\noindent
\textbf{Impedance Leakage at the Gate-level.}

\noindent
\textbf{Impedance Leakage at the Circuit-level.}
\end{comment} 
Impedance side channels, in contrast to EM and power side channels, emerge from changes in impedance caused by switching circuits within a complementary metal–oxide semiconductor (CMOS) device. This unique side channel presents a modulated signal of internal device events, wherein the two-state impedance of transistors in the CMOS device undergoes changes during operation. Our previous works \cite{10133318,PAINE_Paper} represent a mathematical formulation of this phenomenon in \textit{transistor-level}. 
In this work, we consider a two-input CMOS NAND gate as of Figure~\ref{fig:nand_eq_single} to present the idea of information leakage through impedance in \textit{gate-level}. %The output terminal of the NAND gate, denoted as Y, connects to either impedance Zp or Zn, depending on the gate input X. 
% The connection between the CMOS inputs and the measured impedance between source voltage node VCC and ground voltage node GND causes impedance variations during operation. Notably, similar input/output impedance shifting characteristics due to switching are observed in other logic circuits within the CPU. Thus, impedance analysis of CMOS circuits and other logic components serves as an insightful approach to uncovering valuable information through the impedance side channel. 
We consider the NMOS and PMOS to have widths of 2 and 3, respectively, and $C_0$ denotes the fanout capacitance. Depending on the input $A$ and $B$, the metal oxide semiconductor field effect transistors (MOSFETs) can be in the active or cutoff region. Let $Z_{3c}$ be the impedance of $3C$ capacitance and $Z_{eq,0}$ be the impedance of the combined capacitance of $7C$ and $C_{0}$. 
    
%\begin{figure}[htbp]
%    \centering
%    \includegraphics[scale=0.40]{Figures/nand_eq.png}
%    \caption{(a) two-input NAND gate, (b) truth table with relative impedance}
%    \label{fig:nand_eq}
%\end{figure}

% Case 1: Device ON, cut-off
In the cutoff mode, a MOSFET functions as an open switch with very high resistance between the source and drain. Despite this state, a low leakage current exists due to the presence of reverse-biased P-N junctions between the source and drain. The impedance in cutoff mode is dependent on specific device specifications and operating conditions, which can be calculated by measuring the leakage current ($I_{DS,L}$) while considering the drain-to-source voltage ($V_{DS}$).
The leakage current $I_{DS,L}$ has an expression of \cite{baker2019cmos}: 
\begin{align}
    I_{DS,L} = W.C.B^{-B.L_G} \exp{\left( \frac{\Delta \Phi}{V_t} \right)} \left[\exp{\left( \frac{A.V_{DS}}{V_t} \right)} -1 \right]
\end{align} 

Thus, the cutoff region impedance, $Z_C$, is expressed as: 
\begin{align}
    Z_C &= \frac{ V_{DS} }{ I_{DS,L} }   \notag \\
        &= \frac{V_{DS}}{W.C.B^{-B.L_G} \exp{\left( \frac{\Delta \Phi}{V_t} \right)} \left[\exp{\left( \frac{A.V_{DS}}{V_t} \right)} -1 \right]} 
    \label{eq:imp_cutoff}
\end{align}

% Case 2: Device ON, linear/saturation 
In the linear and saturation mode, let $R_{lin,p/n}$ and $R_{sat,p/n}$ be the effective on-resistance % in the linear and saturation regions 
for the PMOS/NMOS, respectively. The expression of $R_{lin,p/n}$ and $R_{sat,p/n}$ can be represented as~\cite{FPGABasedSystemDesign}, 
\begin{align}
    R_{lin,p/n} &= \frac{ \frac{1}{2} (V_{D} - V_{S} - V_{t,p/n}) } {\frac{3}{8} k'_{p/n} (\frac{W}{L})_{p/n} (V_{D} - V_{S} - V_{t,p/n})^{2}} \label{eq:R_lin} \\
    R_{sat,p/n} &= \frac{V_{D} - V_{S}}{\frac{1}{2} k'_{p/n} (\frac{W}{L})_{p/n} (V_{D} - V_{S} - V_{t,p/n})^{2} } \label{eq:R_sat}
\end{align}

Here, $k'_{p/n}$ represents the trans-conductance, $(W/L)_{p/n}$ represents the aspect ratio, and $V_{t,p/n}$ represents the threshold voltage of the PMOS/NMOS. The voltage at the drain is denoted by $V_{D}$, while the voltage at the source is denoted by $V_{S}$. 
Let $R_{p/n}$ be the effective on-resistance of the PMOS/NMOS. 
Using \eqref{eq:R_lin} and \eqref{eq:R_sat}, $R_{p}$ and $R_{n}$ can be estimated as follows: 
\begin{align}
    R_{p/n} = \frac{1}{2} (R_{lin,p/n} + R_{sat,p/n}) \label{eq:R_p/n}
\end{align} 

%\begin{figure}[htbp]
%\centering
%	\includegraphics[scale=0.27]{Figures/mosfet_para_cap.png}
%	\caption{PMOS/NMOS circuit with the parasitic capacitances.}
%	\label{fig:mosfet_para_cap}
%\end{figure}
Fig.~\ref{fig:mosfet_para_cap_single} depicts an equivalent circuit of PMOS/NMOS containing parasitic capacitances. Based on the input $X$ at the gate, the MOSFET changes its operating state from cut-off to saturation following the linear state. 
Let $C_{gdp/n}$ represent the gate to drain capacitance, $C_{dbp/n}$ represent the drain to bulk parasitic capacitance (the diffusion capacitance) for the PMOS/NMOS %, and $C_Y$ represent the capacitance at the output wire 
\cite{DigitalIntegratedCircuits, sedra_smith_2004_digital}. 
% The expressions of the capacitance are presented in Table~\ref{tab:cap_eq}. The parameters associated with Table~\ref{tab:cap_eq} and their definitions are presented in Table~\ref{tab:cap_def}. 
The equivalent capacitance value, $C_{eq,p/n}$, aggregating the parasitic capacitance for the PMOS/NMOS, respectively \cite{FPGABasedSystemDesign} can be expressed as, 
\begin{align}
    C_{eq,p/n} = \sum (C_{gdp/n}, C_{dbp/n}) \label{eq:c_eqp/n}
\end{align} 
Let $X_{p/n}$ be the equivalent reactance of NMOS/PMOS, respectively. Interestingly, the parasitic capacitance of the PMOS/NMOS dominates the equivalent reactance $X_{p/n}$ \cite{DigitalIntegratedCircuits}. Hence, the equivalent reactance $X_{p/n}$ follows,  
\begin{align}
    X_{p/n} \approx \frac{-1}{\omega C_{eq,p/n}} \label{eq:X_p/n}
\end{align}
Here,  $\omega = 2\pi f$ and $f$ is the signal frequency. % and the total equivalent capacitance of the PMOS/NMOS between the output node $Y$ and the ground node $GND$ is $C_{eq}$. 
The equivalent impedance, $Z_{p/n}$ for PMOS/NMOS, consists of equivalent resistance $R_{p/n}$ and equivalent reactance $X_{p/n}$. The expression of the equivalent impedance $Z_{p/n}$ can be expressed as,   
\begin{align}
    Z_{p/n} = R_{p/n} + j X_{p/n} \label{eq:active_Z}
\end{align} 
Using \eqref{eq:active_Z}, we can calculate the impedance for the PMOS and NMOS working in the linear and saturation regions. The physical parameters for the PMOS and NMOS are distinct. Therefore, the impedance values for the PMOS and NMOS operating in the linear and saturation regions are not identical.

%  two-input NAND gate impedance
To demonstrate how information is leaked through impedance in the \textit{gate-level}, we consider a simple two-input NAND gate (Fig.~\ref{fig:nand_eq_single}). Let $Z_{P,a}$, $Z_{N,a}$ be the active region impedance and $Z_{P,c}$, $Z_{N,c}$  be the cutoff region impedance for the PMOS and NMOS of the NAND gate, respectively. Equation~\eqref{eq:active_Z} can be used to estimate $Z_{P,a}$ and $Z_{N,a}$. Equation~\eqref{eq:imp_cutoff}, on the other hand, can be used to estimate $Z_{P,c}$ and $Z_{N,c}$. Using $Z_{P/N,a/c}$, the impedance values for all four possible input patterns to the two-input NAND gate can be estimated. 

\textbf{Case $A=0$, $B=0$:} Here, the two PMOS remain active while the two NMOS remain inactive. Let, the impedance between $V_{DD}$ and ground node be $Z_{00}$ and it can be expressed as, 
\begin{align}
    Z_{00}  &= Z_{P,a} || Z_{P,a} + (Z_{N,c} || Z_{3C} + Z_{N,c}) || Z_{eq,0} \nonumber \\
            &= \frac{Z_{P,a}}{2} + \frac{(Z_{N,c} Z_{3C} + Z_{N,c} (Z_{N,c} + Z_{3c})) Z_{eq,0} }{  Z_{N,c} Z_{3C} + (Z_{N,c} + Z_{eq,0}) (Z_{N,c} + Z_{3c}) } \label{eq:z00}
\end{align}

Similarly,  for \textbf{Case $A=0$, $B=1$}, \textbf{Case $A=1$, $B=0$}, and \textbf{Case $A=1$, $B=1$}, the impedance from the supply voltage side can be expressed in terms of $Z_{01}$ \eqref{eq:z01}, $Z_{10}$ \eqref{eq:z10}, and $Z_{11}$ \eqref{eq:z11}, respectively.  

\begin{align}
    Z_{01}  &= Z_{P,a} || Z_{P,c} + (Z_{N,c} || Z_{3C} + Z_{N,a}) || Z_{eq,0} \hspace{10pt}  \label{eq:z01} \\
    Z_{10}  &= Z_{P,c} || Z_{P,a} + (Z_{N,a} || Z_{3C} + Z_{N,c}) || Z_{eq,0} \hspace{10pt}  \label{eq:z10} \\
    Z_{11}  &= Z_{P,c} || Z_{P,c} + (Z_{N,a} || Z_{3C} + Z_{N,a}) || Z_{eq,0} \hspace{10pt}  \label{eq:z11}
\end{align}

This section considers a simple two-input NAND gate to demonstrate the concept of different input combinations producing distinct impedance values. The equivalent impedance between $V_{DD}$ and ground nodes as expressed by \eqref{eq:z00}-\eqref{eq:z11} are distinct for all the possible input combinations of $A$ and $B$. Although this is a simple scenario, a digital system contains different types of logic gates to carry out different types of functions. %In our future study, we will investigate how inputs influence the impedance between the voltage source and ground of other logic gates. We can utilize this to estimate device level impedance which can further be helpful in extracting the uncontrollable inputs of a device. We will also investigate the signal strength of the distinct impedance.   

\section{Exploiting Impedance Leakages to Reverse-engineer Embedded Software} 
\label{sec:ILinReverseEngineering}
\noindent 
%% briefly introduce impedace side channel. if possible explain briefly how it is different than potential side channels, i.e., power, em side channel. 
%% talk about the potential of impedance side channel in various fields. i.e., detecting counterfeit hardware/pcb, extracting encryption keys, even detecting program instructions being executed inside a cpu. then end it by stating that in our previous work we had explored how impedance can be used to detect program instruction "types" and in this paper our goal is to reverse engineer software instructions .
% Traditionally electronic device impedance is regarded as a static and unassuming property but it has potential as an invaluable source of information leakage as described in Section~\ref{sec:imp_model}. 
Unlike conventional side channels such as power and electromagnetic (EM) emissions that monitor fluctuations in power consumption or EM radiation, impedance side channels directly characterize the intrinsic impedance of a device across a range of frequencies. 
This novel perspective of impedance side channel transcends the conventional understanding of impedance, which typically focuses on its role in circuit design and signal transmission. Instead, the impedance side channel considers impedance as a dynamic and evolving parameter that can inadvertently leak sensitive information about the internal operations of embedded systems Section~\ref{sec:imp_model}. 
% The device impedance can reveal crucial insights into the internal operations of a device \cite{}. 
Consequently, the impedance side channel can be utilized in detecting counterfeit hardware and printed circuit boards (PCBs) by profiling authentic components and identifying impedance mismatches \cite{zhang2015robust,nguyen2019creating}. Moreover, variations in device impedance during cryptographic operations can be exploited to extract secret encryption keys \cite{9191655}. Remarkably, even the execution of different program instructions induces discernible impedance changes that can be leveraged to monitor software behavior \cite{10133318}. 
% One crucial aspect lies in verifying the integrity of executed software instructions. The software instructions are processed sequentially within an embedded device, and a skilled attacker can target them as they transfer from non-volatile memory to registers. In this context, monitoring executed software instructions offers a means to compute the integrity of their immutable portions and compare them against predetermined values. 

Our previous work \cite{10133318} demonstrates the potential of impedance side channel for classifying program instruction types, such as data transfer, arithmetic, logical, and branch instructions. Distinct impedance profiles are observed during the execution of each instruction type. 
Building on these insights, this paper explores the use of device impedance as a side channel specifically for reverse engineering individual software instructions. By observing device impedance variations during software instruction execution, it may be possible to identify the exact sequence of instructions being run on a device. Moreover, reverse engineering software instructions via impedance side channels can enable numerous hardware security applications. It can be used to validate the integrity of program execution or detect the presence of malicious instructions, such as those injected through malware. Additionally, identifying instruction sequences can help expose counterfeit systems lacking developer fingerprints. Overall, impedance side channel analysis promises to unlock new perspectives on the vulnerability landscape of embedded systems \cite{10133318}.

% Our motivation in this research is to explore the feasibility of utilizing device impedance as an emerging side channel for the disassembly of software instructions—a novel and promising approach that bridges the gap between hardware characteristics and software security. 
The motivation of this research is to evaluate the reliability and potential of adopting impedance side channel analysis, a hardware-based approach, for disassembling individual software instructions. By transforming impedance from a static, passive property into an active information source, we can bridge the gap between hardware characteristics and software security. 
The ability to link impedance profiles to instruction behavior can revolutionize reverse engineering and greatly empower the protection of connected systems against evolving cyber threats \cite{omar2022defending,wang2022rethinking}.

\section{Experimental Setup}
\label{sec:experimental_setup} \noindent 
The experiment proceeds in three steps. In this section, we describe the physical hardware setup, detailing the equipment configuration. Next, the software code and instructions executed are outlined. Finally, the signal acquisition process is elucidated, explaining how data is collected. 

\subsection{Hardware Setup} \noindent 
Our goal in this experiment is to observe if impedance side channel can help reverse engineer software instructions. We also to study the impact of hardware complexity on this new side channel. To do this, we conduct experiments on two different platforms: the Arduino Uno~\cite{arduinoUNO}, which uses an 8-bit AVR ATmega328P microcontroller, and the Alchitry AU with an Artix 7 FPGA~\cite{alchitry}. We choose these platforms due to their differing levels of complexity and configurability. 
The main differences between microcontrollers like the Arduino and FPGAs like the Alchitry are: 
\begin{itemize}
    \item FPGAs can process multiple inputs in parallel, while microcontrollers execute instructions one at a time sequentially. 
    \item FPGA hardware can be reconfigured and customized, but microcontroller hardware has limited flexibility. 
\end{itemize}

\begin{figure}[htbp]
    \centering
    \includegraphics[width= 0.480\textwidth]{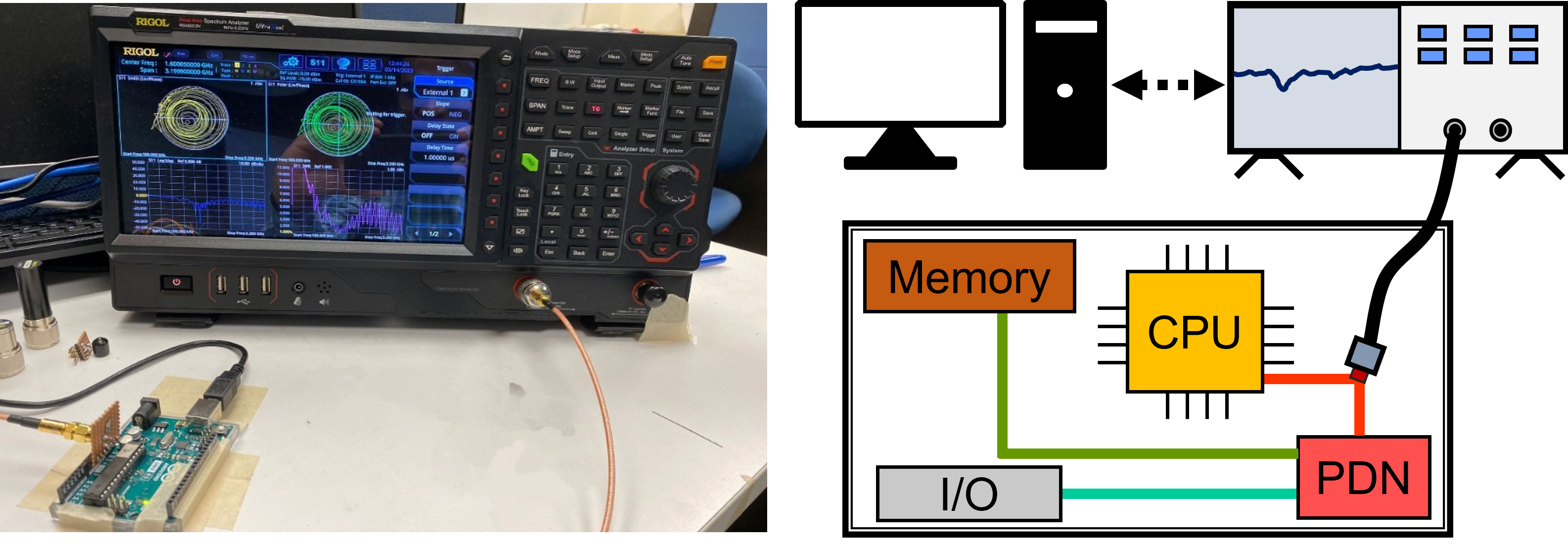}
    \caption{Hardware setup: measurement of \textit{impedance} signals.}
    \label{fig:experimental_setup}
\end{figure}
Figure~\ref{fig:experimental_setup} illustrates the hardware setup for our experiments on the Arduino and Alchitry systems. 
For both devices, we target the 3.3V power domain since it is closer to the central processing unit (CPU) voltage level. The Arduino and Alchitry both have onboard 3.3V pins, so no physical modification or tampering is required to tap into this domain and monitor impedance changes near the core of each device. 
By leveraging the 3.3V pins, we observe the impedance side channel on both platforms non-invasively. This highlights a key benefit of impedance side channel—the ability to monitor signals without direct hardware access or modification. The use of native 3.3V pins enables aboveboard measurements, avoiding tampering. 

We use a Rigol RSA5032N spectrum analyzer in vector network analyzer (VNA) mode to measure impedance. We connect VNA to each device's power distribution network (PDN) using a coaxial cable. The coaxial cable consists of an inner conductor surrounded by a concentric conducting shell, which enables signal transmission in the center while shielding against interference. The cable's design minimizes signal loss and electromagnetic interference (EMI). We measure impedance fluctuations with better precision by directly attaching the VNA to the PDN. The VNA-to-PDN connection enables us an accurate means to monitor impedance changes related to processor activity in each device. 
%Here we use a special experimental setup to let the impedance traces vary slightly across runs. 
%Therefore, EMI from the environment is not an issue in our experimental setup. 
%Thus, we observe low variation across runs and a similar trend in the curves for the same instance.

\begin{table*}[htbp]
\renewcommand{\arraystretch}{1}
\caption{Program instruction set summary.}
\label{tab:ins_summary}
\centering
\begin{tabular}{|p{4.5em}|p{12.5em}||p{3em}|p{5em}|p{9em}||p{3em}|p{5em}|p{9em}|}
\hline    

\multirow{2}{4em}{\bfseries Instruction type} & \multirow{2}{5em}{\bfseries Description} 
& \multicolumn{3}{c||}{\bfseries Artix 7 FPGA} & \multicolumn{3}{c|}{\bfseries ATmega328P microcontroller} \\ \cline{3-8} 
& & \bfseries Opcode & \bfseries Operands & \bfseries Operation & \bfseries Opcode & \bfseries Operands & \bfseries Operation \\ \hline

  \multirow{3}{5em}{Data transfer} 
  & Load between registers        & LOAD  & R1, R2, K &  R1 $\leftarrow$ [R2 $+$ K]      & MOV &  R1, R2 & R1 $\leftarrow$ R2  \\ \cline{2-8}
  & Store from register to memory       & STORE & R1, R2, K &  R1 $\rightarrow$ [R2 $+$ K]     & --- & ---        &  ---                         \\ \cline{2-8}
  & Load constant into register       & SET   & R1, K     &  R1 $\leftarrow$ K               & LDI &  R1, K     & R1 $\leftarrow$ K           \\ \hline 
  
  \multirow{5}{5.5em}{Arithmetic and logic} 
  & Addition                     & ADD & R1, R2, R3  &  R1 $\leftarrow$ R2 $+$ R3      & ADD & R1, R2  & R1 $\leftarrow$ R1 $+$ R2       \\ \cline{2-8}
  & Subtraction                  & SUB & R1, R2, R3  &  R1 $\leftarrow$ R2 $-$ R3      & SUB & R1, R2  & R1 $\leftarrow$ R1 $-$ R2       \\ \cline{2-8}
  & Bit-wise AND               & AND & R1, R2, R3  &  R1 $\leftarrow$ R2 $\&$ R3     & AND & R1, R2  & R1 $\leftarrow$ R1 $\&$ R2      \\ \cline{2-8}
  & Bit-wise OR                & OR  & R1, R2, R3  &  R1 $\leftarrow$ R2 $|$ R3      & OR  & R1, R2  & R1 $\leftarrow$ R1 $|$ R2       \\ \cline{2-8}
  & Bit-wise XOR               & XOR & R1, R2, R3  &  R1 $\leftarrow$ R2 $\oplus$ R3 & EOR & R1, R2  & R1 $\leftarrow$ R1 $\oplus$ R2  \\ \hline 
  
  \multirow{2}{5.5em}{Rotate}
  & Binary left shift          & SHL & R1, R2, R3  &  R1 $\leftarrow$ R2 $<<$ R3     & LSL & R1      & R1[n+1] $\leftarrow$ R1[n] \newline R[0] $\leftarrow$ 0     \\ \cline{2-8} 
  & Binary right shift         & SHR & R1, R2, R3  &  R1 $\leftarrow$ R2 $>>$ R3     & LSR & R1      & R1[n]   $\leftarrow$ R1[n+1] \newline R[7] $\leftarrow$ 0   \\ \hline 

  \multirow{2}{5.5em}{Branch}
  & Branch if equal      & BEQ  & R1, K  & if R1 $=$ K: \newline PC $\leftarrow$ PC $+$ 2    & BREQ & K & if Z $=$ 1: \newline PC $\leftarrow$ PC$+$K$+$ 1  \\ \cline{2-8}      
  & Branch if not equal  & BNEQ & R1, K  & if R1: $\neq$ K \newline PC $\leftarrow$ PC $+$ 2 & BRNE & K & if Z $=$ 0: \newline PC $\leftarrow$ PC$+$K$+$1  \\ \hline 
\end{tabular}
\end{table*}

\subsection{Software Setup} \noindent 
We program the test devices using assembly language, which is a low-level programming language with direct correspondence to hardware machine code. 
Assembly instructions consist of an opcode mnemonic followed by operands like data, arguments, or parameters. The opcode indicates the operation to be performed, while the operands provide the associated data. Assembly language allows us to write code with the CPU's basic instructions. This level of control enables us to understand the fundamental operations occurring during code execution. 

For the Artix 7 FPGA experiments, we construct a simple 8-bit CPU with a custom instruction set. We also develop a custom assembler to translate the instructions into memory writes for the FPGA. 
In total, we define 12 instruction types: data transfer (LOAD, STORE, SET), arithmetic/logic (ADD, SUB, AND, OR, XOR), rotate (SHL, SHR), and branch (BEQ, BNEQ). This custom 12-instruction set enables basic program execution on our FPGA CPU for testing the impedance side channel. Table~\ref{tab:ins_summary} summarizes the instruction operations, which are briefly described below. 
% The data transfer instructions move data between registers and memory. The arithmetic/logic instructions perform basic mathematical and boolean operations. The rotate instructions shift the bits left or right. Finally, the branch instructions implement conditional jumps depending on a comparison. 
\begin{description}
    \item[\textit{LOAD, STORE, SET:}] \hspace{65pt} LOAD and SET transfer data into registers, while STORE transfers data from a register to memory. 
    LOAD and STORE use the syntax: \textit{Opcode R1, R2, K}. For LOAD, R1 receives the value. For STORE, R1 provides the output value. R2 contains the base address, and K is an offset constant. 
    SET uses the syntax: \textit{Opcode R1, K}. R1 is the target register, and K is the 8-bit value to assign.
    
    \item[\textit{ADD, SUB, AND, OR, XOR:}] \hspace{95pt} The arithmetic and logic instructions (ADD, SUB, AND, OR, XOR) follow the syntax: \textit{Opcode R1, R2, R3}. They perform operations between registers R2 and R3, storing the result in R1. 
    Specifically, ADD does addition, SUB does subtraction, AND performs a logical AND, OR does a logical OR, and XOR executes an exclusive OR. 

    \item[\textit{SHL, SHR:}] \hspace{20pt} These two instructions have the syntax: \mbox{\textit{Opcode R1, R2, R3}}. The SHL and SHR instructions shift the bits in the R2 register value left or right, respectively, by the number of bit positions specified in the R3 register and store the result  in the R1 register. The least significant bits (LSBs) and most significant bits (MSBs) are padded with zeros following the SHL and SHR, respectively. 
    
    \item[\textit{BEQ, BNEQ:}] \hspace{30pt} The BEQ and BNEQ instructions follow the syntax: \mbox{\textit{Opcode R1, K}} and allow conditional branching based on comparing a register, R1, and constant, K. 
    BEQ skips the next instruction by increasing programming counter, $PC$, by 2 if they are equal. BNEQ does the opposite (skips is they are not equal).
\end{description}

\vspace{2pt}
Similarly, for the Arduino's ATmega328P AVR microcontroller, we adhere to the conventional assembly instructions provided by the manufacturer~\cite{atmegadatasheet}. We select 11 representative instructions, summarized in Table~\ref{tab:ins_summary}, consisting data transfer, arithmetic/logic, rotate (SHL, SHR), and branch operations. The 8-bit ATmega328P is a RISC microcontroller with 1 KB EEPROM, 2 KB SRAM, 23 GPIO, and 32 general registers. 
The chosen instruction set mirrors the functionality of the custom FPGA instructions, leveraging the ATmega328P's standard feature set. The following paragraph briefly outlines the operations of the instructions. 
\begin{description}
    \item[\textit{MOV, LDI:}] \hspace{20pt} MOV instruction is used to transfer data from one register to another, while LDI instruction is used to load an immediate to the  designated register.  
    
    \item[\textit{ADD, SUB, AND, OR, EOR:}] \hspace{95pt} These instructions execute the arithmetic and logical operations. ADD, SUB, AND, OR, EOR perform the addition, subtraction, logical AND, OR, and XOR operations, respectively. 

    \item[\textit{LSL, LSR:}] \hspace{20pt} LSL, and LSR shift one bit of the specified register value to the left or right, respectively. The LSBs and MSBs are padded with zeros following the LSL and LSR instructions, respectively. 
    
    \item[\textit{BREQ, BRNE:}] \hspace{30pt} The BREQ (branch if equal) and BNEQ (branch if not equal) instructions control program flow based on the zero flag, $Z$. They are often used after arithmetic operations that set or reset $Z$. If $Z$ is 0, BREQ will branch to a different instruction sequence instead of the next immediate instruction. BNEQ branches if $Z$ is not 0. By manipulating the program counter ($PC$), these instructions create conditional loops and branching logic. 
\end{description}

\begin{figure}[htbp]
\centering
\begin{subfigure}{.23\textwidth}
    \raggedright
    \includegraphics[width = 0.95\textwidth]{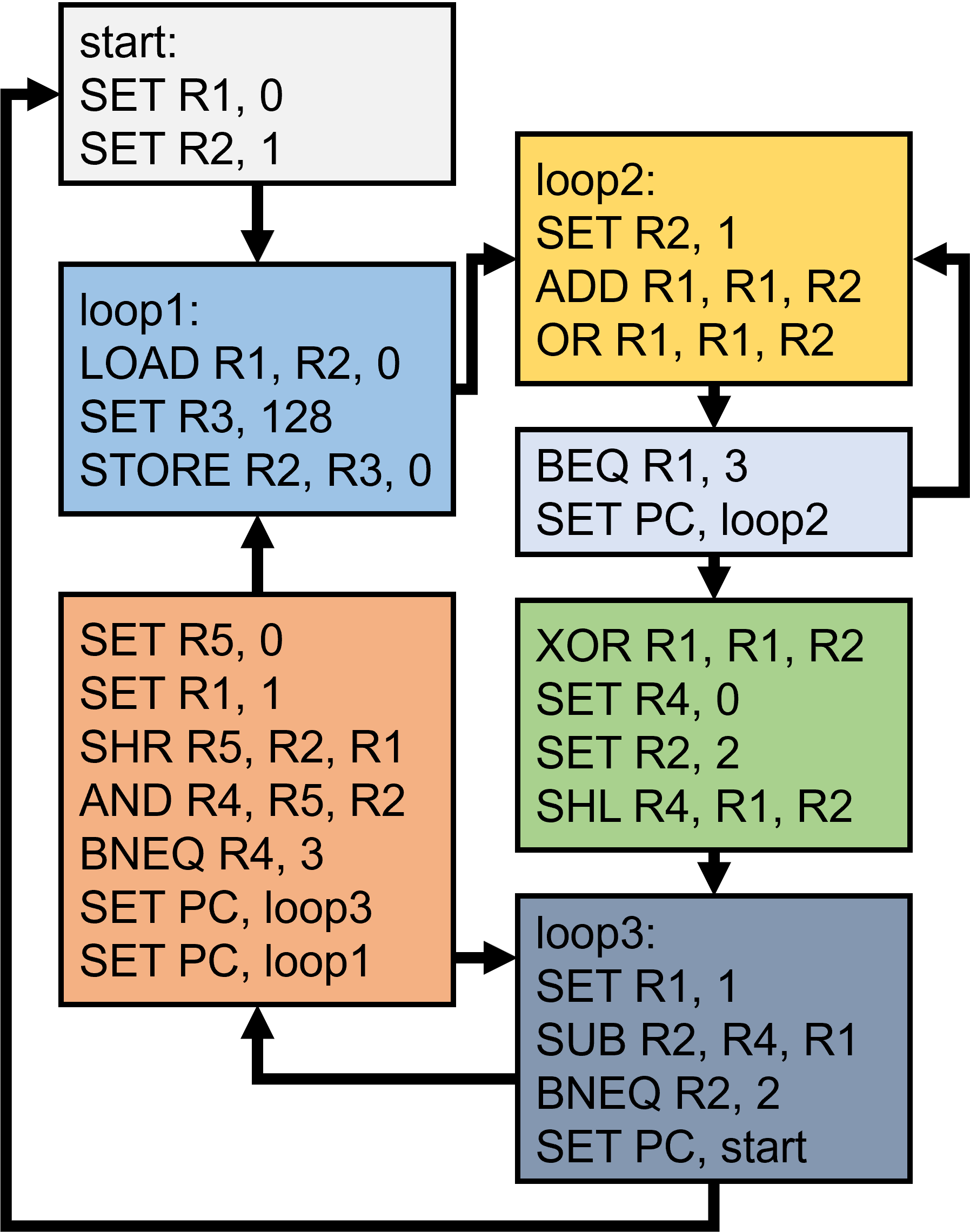}
    \caption{}
    \label{fig:program_stepsFPGA}
\end{subfigure}%
\begin{subfigure}{.23\textwidth}
    \raggedleft
    \includegraphics[width = 0.95\textwidth]{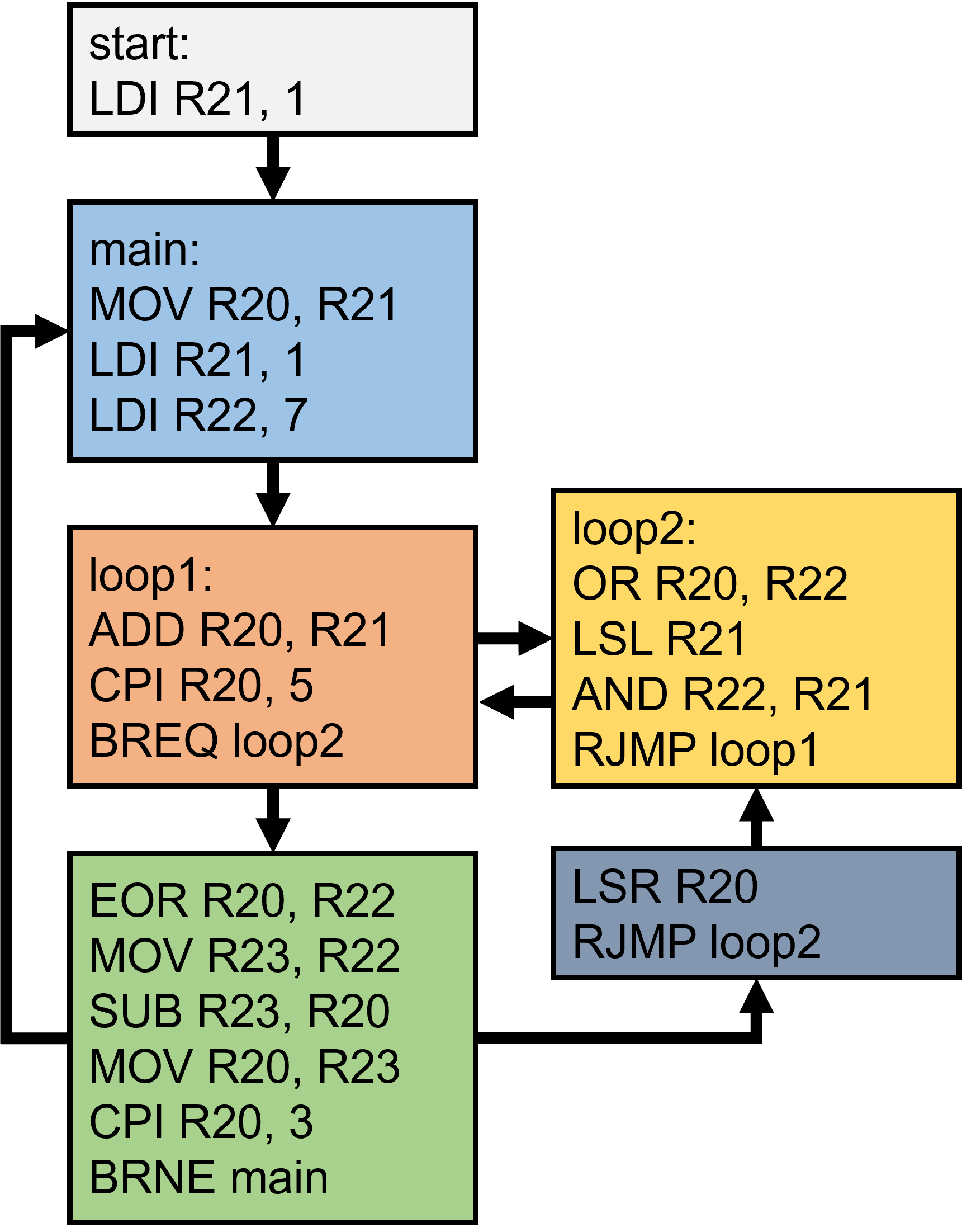}
    \caption{}
    \label{fig:program_stepsATmega}
\end{subfigure}
\caption{Assembly code snippet followed in the experiment: (a) Artix 7 FPGA, (b) ATmega328P microcontroller.}
\label{fig:code_snippet}
\end{figure}

The assembly code executed for the experiment runs all instructions in Table~\ref{tab:ins_summary} on both the Artix 7 FPGA and ATmega328P microcontroller platforms. Figure~\ref{fig:code_snippet} and Figure~\ref{fig:program_stepsATmega} illustrate simplified block diagrams representing the instruction flow for the FPGA and microcontroller, respectively. We follow these instruction flows to collect impedance signals with no particular computational purpose. 
Figure~\ref{fig:program_stepsATmega} includes two additional instructions, \textit{CPI} and \textit{RJMP}, used only for the ATmega328P to control program flow. The \textit{CPI} instruction compares an immediate value to a register value and sets the zero flag $Z$. The \textit{RJMP} instruction performs a relative jump to a specified location in the code. % In summary, the same core instruction set is executed on both platforms, with minimal additional flow control for the microcontroller. The instruction sequences serve purely to generate impedance profiles for each operation.

% assembly instructions (table), nstruction counter
\subsection{Signal Collection} \noindent 
In our study, we use the Rigol RSA5032N spectrum analyzer configured as a vector network analyzer (VNA). The RSA5032N contains built-in circuitry to perform VNA measurements and supports standard commands for programmable instruments (SCPI). SCPI instructions transfer the trace data over Ethernet in complex format. Therefore, the received complex trace values need to be transformed into impedance values for analysis.

Let $T_{Re,m}$, and $T_{Im,m}$ represent the real and imaginary portions of the trace at frequency $m$. First, we convert the real and imaginary components to resistance $R_m$, and reactance, $X_m$ using Equation.~\ref{eq:TraceResistance} and Equation.~\ref{eq:TraceReactance}, respectively~\cite{dunsmore2020handbook}. All raw traces are collected during software instruction execution and transformed into impedance signals. 
\begin{align}
    R_m &= Z_{ref} * \frac{(1 - T_{Re,m}^2 - T_{Im,m}^2)}{(1 - T_{Re,m})^2 + T_{Im,m}^2}  \label{eq:TraceResistance} \\
    X_m &= Z_{ref} * \frac{ 2*T_{Im,m} }{(1 - T_{Re,m})^2 + T_{Im,m}^2}  \label{eq:TraceReactance}
\end{align}
Here, $Z_{ref}$ is the reference impedance of the measurement instrument. The overall complex impedance, $Z_{T,m}$, at the trace point $m$ is calculated as, 
\begin{align}
    Z_{T,m} = R_m + i X_m \label{eq:Z_trace}
\end{align}

We specify 10,001 linearly spaced frequency points from 500kHz to 4GHz. At each point, we compute the average of 100 sampled measurements. This is repeated to collect signals for each instruction execution. Using a custom Python script with SCPI commands, we collect 700 signals per FPGA instruction (8,400 total) and 500 signals per microcontroller instruction (5,500 total). 
These raw signals are in complex form. The complex trace signals are converted to impedance using Equation~\ref{eq:Z_trace}, where $Z_{ref}$ is the 50 $\Omega$ reference impedance. By transforming the complex trace values into impedance, we obtain the necessary data for analysis. The resulting impedance signals are then labeled for supervised learning. 

\begin{figure}[htbp]
\centering
\begin{subfigure}{.245\textwidth}
  \raggedright
  \includegraphics[width = 1\textwidth]{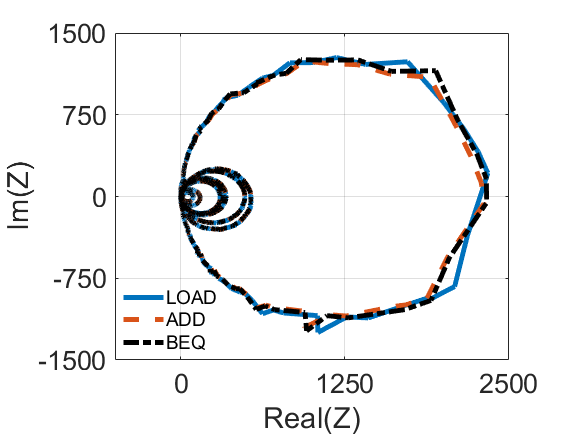}
  \caption{}
  \label{fig:smith_fpga}
\end{subfigure}%
\begin{subfigure}{.245\textwidth}
  \raggedleft
  \includegraphics[width = 1\textwidth]{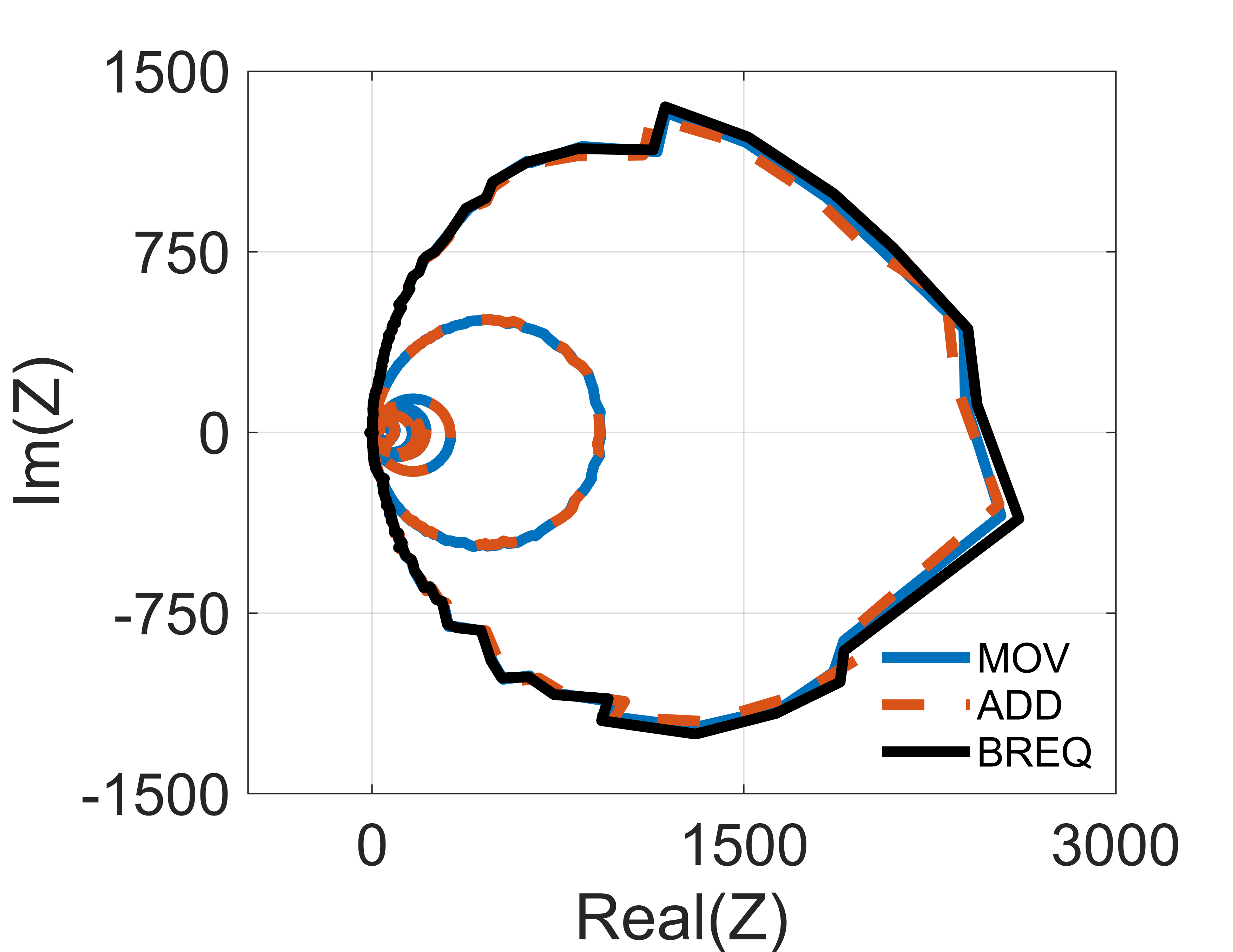}
  \caption{}
  \label{fig:smith_atmega}
\end{subfigure}
\caption{Impedance profile of program instructions. (a) Artix 7: LOAD, ADD, BEQ, (b) ATmega328P: MOV, ADD, BREQ.}
\label{fig:smith}
\end{figure}

\section{Evaluation} 
\label{sec:evaluation} 
\subsection{Signal Analysis} 
\begin{figure}[htbp]
\centering
\begin{subfigure}{.245\textwidth}
  \raggedright
  \includegraphics[width = 1.1\textwidth]{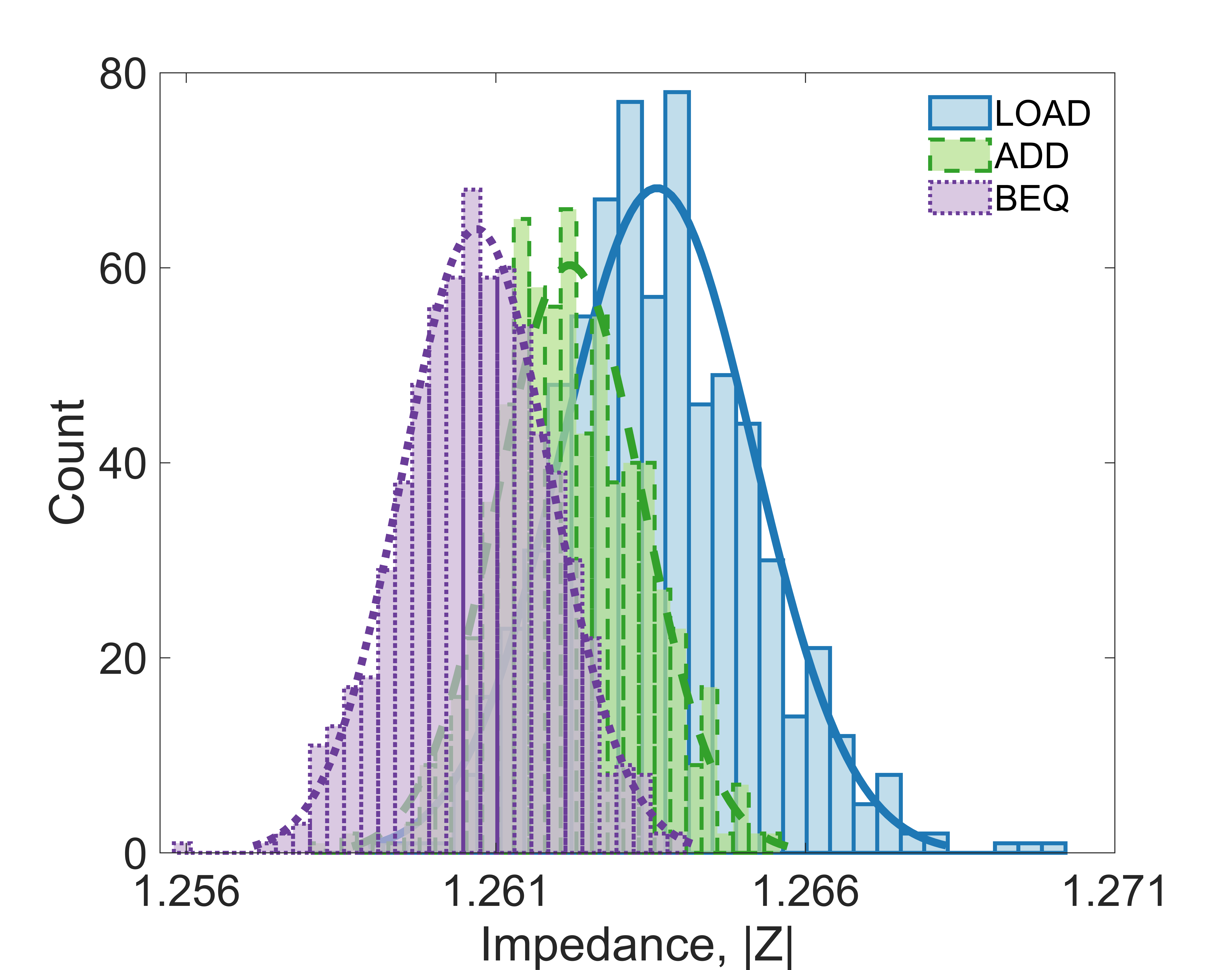}
  \caption{}
  \label{fig:fpga1.2ghz}
\end{subfigure}%
\begin{subfigure}{.245\textwidth}
  \raggedleft
  \includegraphics[width = 1.1\textwidth]{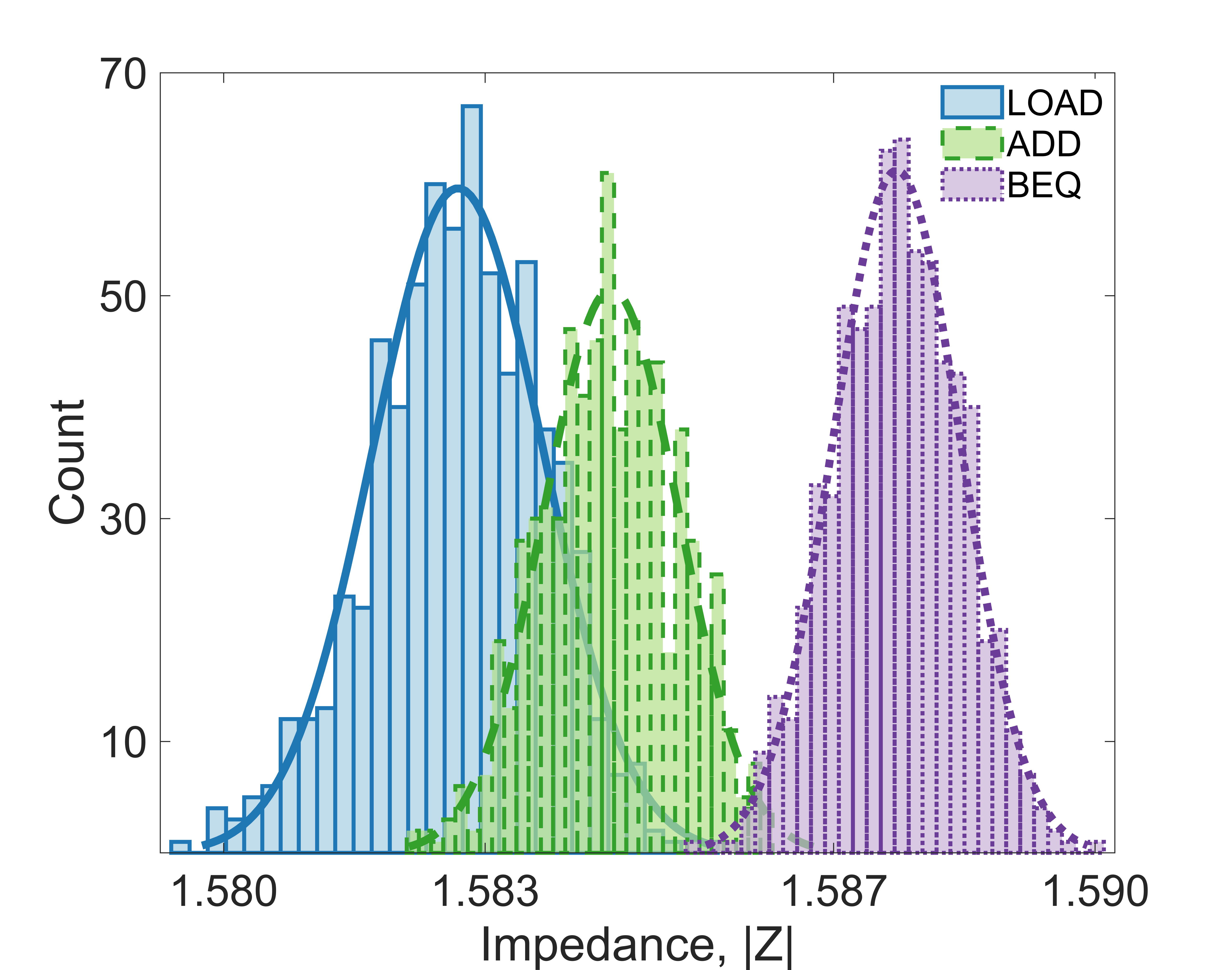}
  \caption{}
  \label{fig:fpga2.3ghz}
\end{subfigure}
\begin{subfigure}{.245\textwidth}
  \raggedright
  \includegraphics[width = 1.1\textwidth]{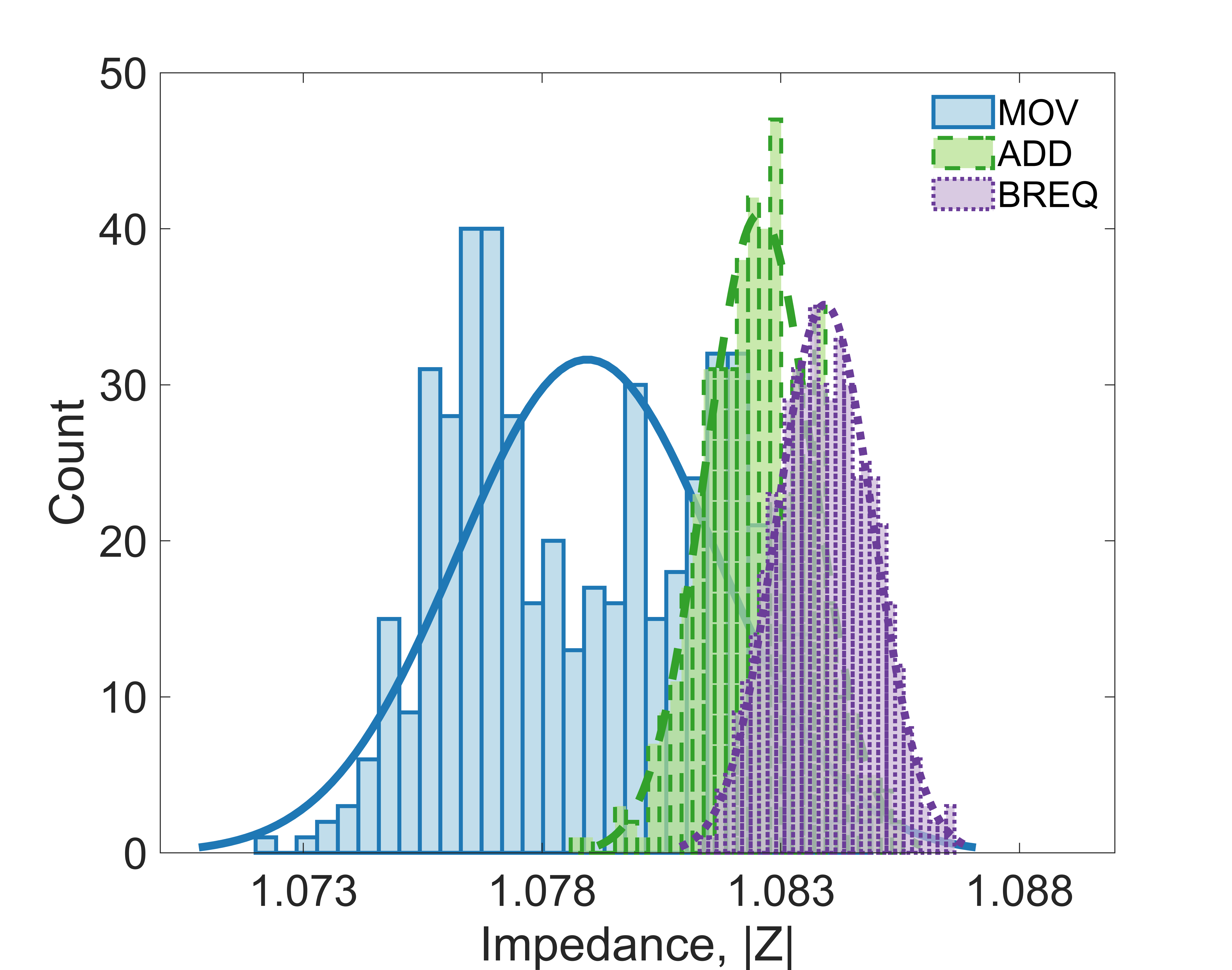}
  \caption{}
  \label{fig:atmega1.2ghz}
\end{subfigure}%
\begin{subfigure}{.245\textwidth}
  \raggedleft
  \includegraphics[width = 1.1\textwidth]{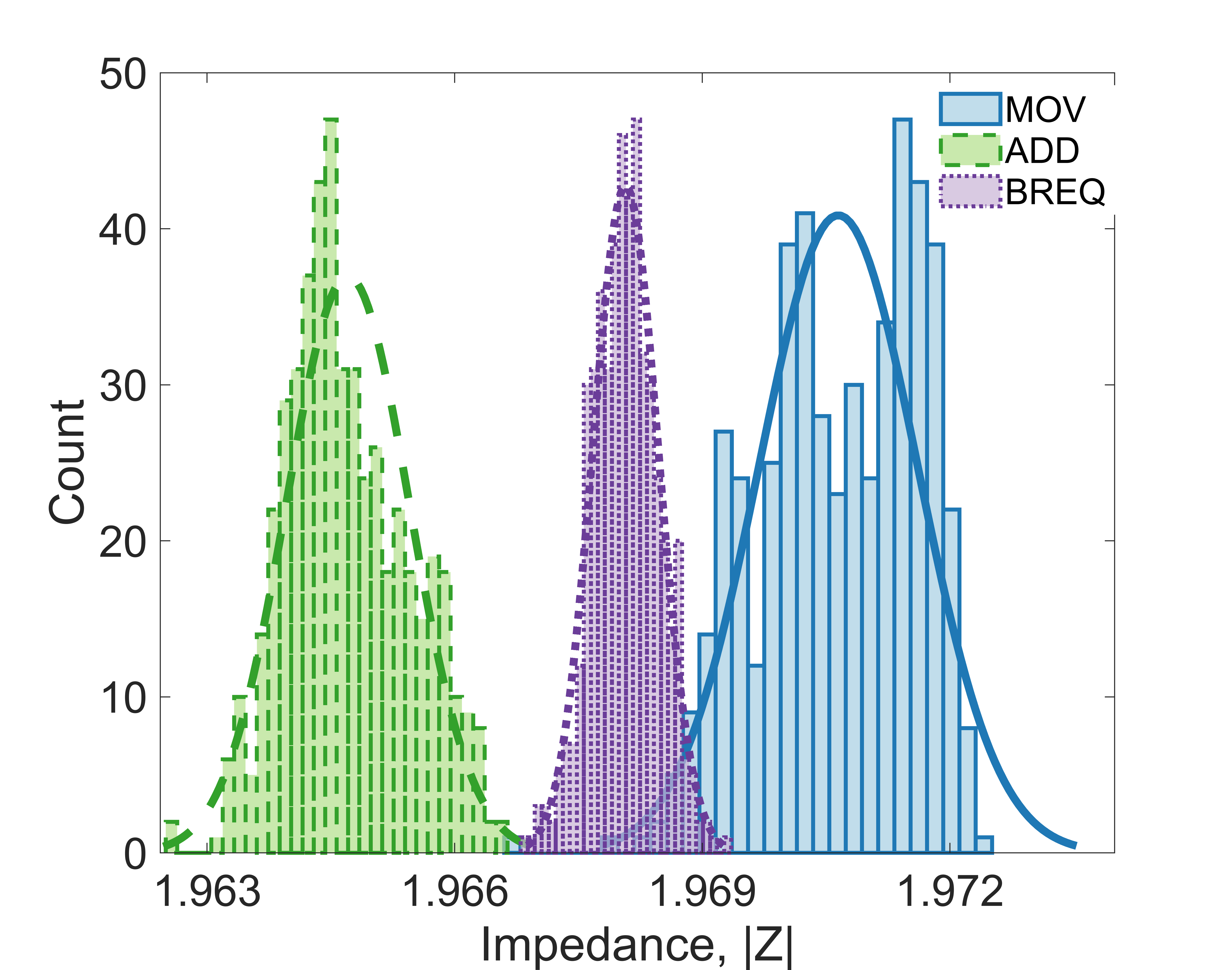}
  \caption{}
  \label{fig:atmga2.3ghz}
\end{subfigure}
\caption{Impedance profile distribution: Artix 7 at (a) 1 GHz, (b) 2.3 GHz, and ATmega328P at (c) 1 GHz, (d) 2.3 GHz.}
\label{fig:distribution}
\end{figure}

\noindent 
The impedance profiles for each instruction execution are visibly distinct, as seen in Figure~\ref{fig:smith_fpga} for three Artix 7 operations (LOAD, AND, BNEQ) and Figure~\ref{fig:smith_atmega} for the ATmega328P (MOV, ADD, BREQ). Though subtle, variations between instructions become evident at certain frequencies. This motivates the removal of redundant frequency points from the raw impedance signals. 
Further analysis uses the impedance magnitude values as they present most of the information~\cite{PAINE_Paper,awal2022nearfield}. At each measured frequency, the signals follow a Gaussian distribution, representing additive white Gaussian thermal noise. Figure~\ref{fig:distribution} histograms illustrate the normal distribution of impedances at two of the selected frequencies. 

\begin{comment}
\begin{figure*}[htbp]
\centering
\begin{subfigure}{.45\textwidth}
  \centering
  \includegraphics[width = 1\textwidth]{Figures/pdf500mhz.eps}
  \caption{Sample distribution at 500 MHz.}
  \label{fig:500mhz}
\end{subfigure}%
\begin{subfigure}{.45\textwidth}
  \centering
  \includegraphics[width = 1\textwidth]{Figures/pdf1ghz.eps}
  \caption{Sample distribution at 1.2 GHz.}
  \label{fig:1.2ghz}
\end{subfigure}
\begin{subfigure}{.45\textwidth}
  \centering
  \includegraphics[width = 1\textwidth]{Figures/pdf2.3ghz.eps}
  \caption{Sample distribution at 2.3 GHz.}
  \label{fig:2.3ghz}
\end{subfigure}
\begin{subfigure}{.45\textwidth}
  \centering
  \includegraphics[width = 1\textwidth]{Figures/pdf3ghz.eps}
  \caption{Sample distribution at 3 GHz.}
  \label{fig:3ghz}
\end{subfigure}
\caption{Distribution of impedance at various frequency points.}
\label{fig:distribution}
\end{figure*}
\end{comment}

\begin{figure*}[!htbp]
\centering
\begin{subfigure}{.49\textwidth}
    \raggedright
    \includegraphics[width = 0.95\textwidth]{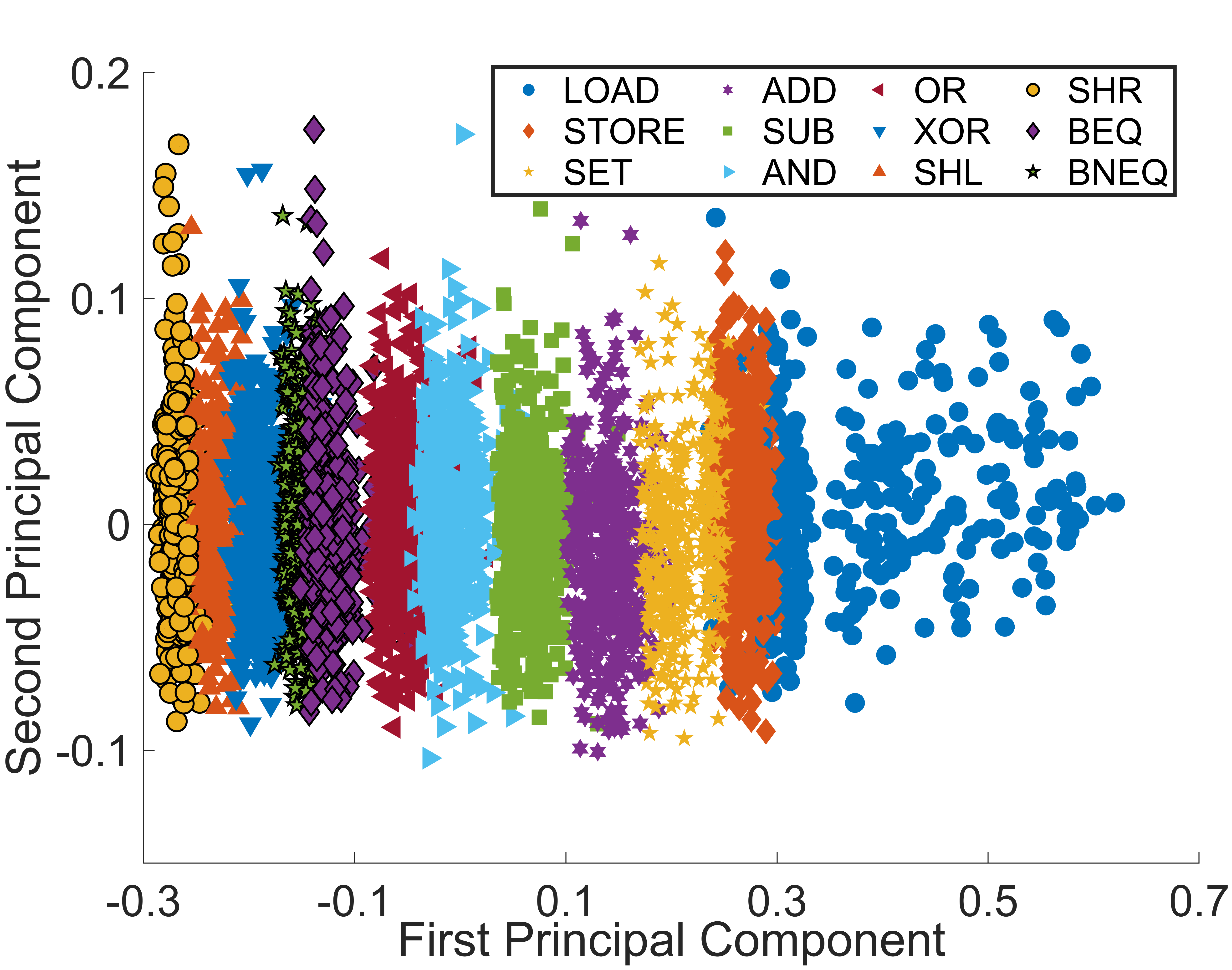}
    \caption{Artix 7: two major principal component distributions.}
    \label{fig:indv_features_fpga}
\end{subfigure}
%\end{figure}
%\begin{figure}
%\centering
\begin{subfigure}{.49\textwidth}
    \raggedleft
    \includegraphics[width = 0.95\textwidth]{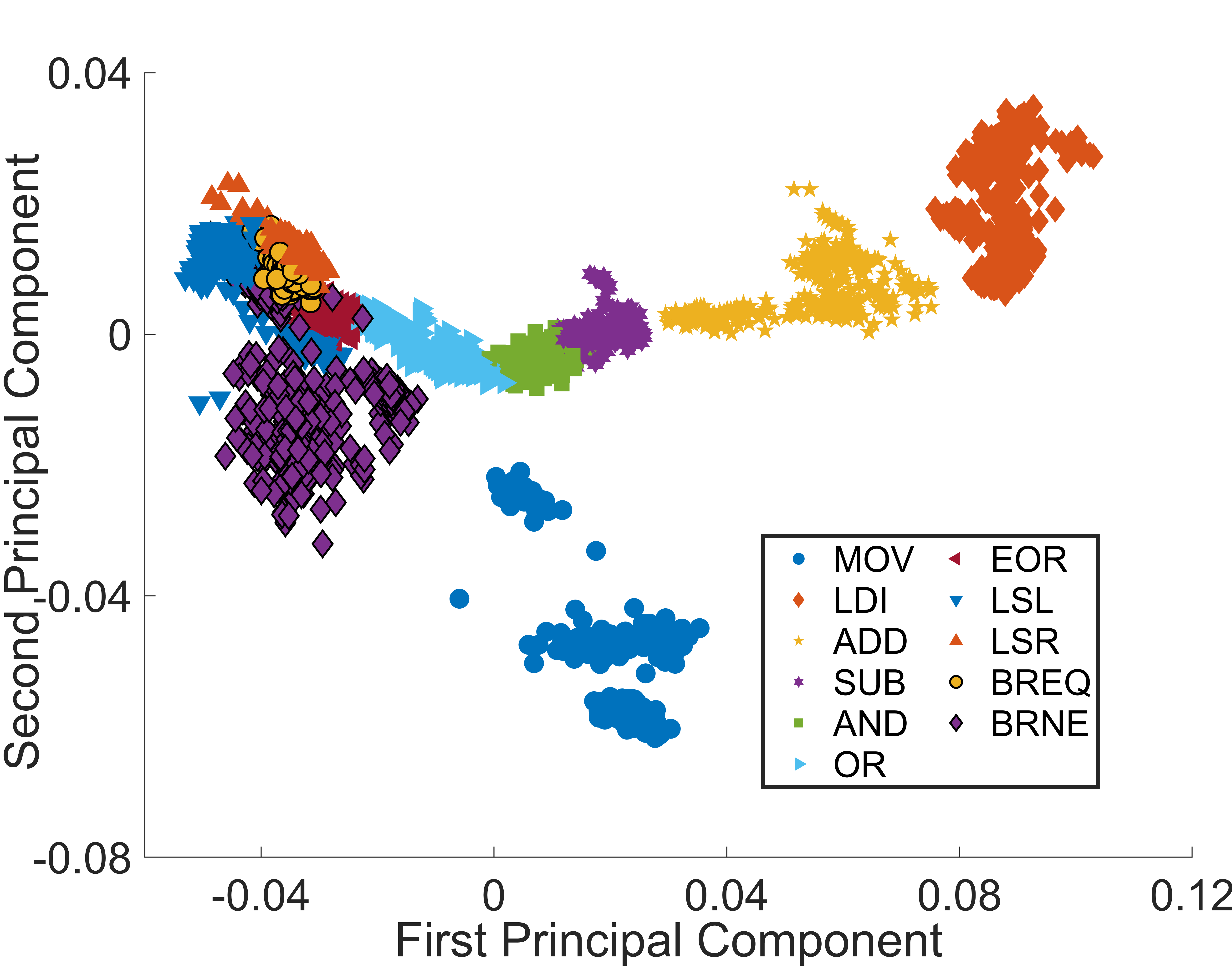}
    \caption{ATmega328P: two major principal component distributions.}
    \label{fig:indv_features_atmega}
\end{subfigure}
\caption{Distribution of the two most significant principal components for multiple program instructions.}
\label{fig:PCAdistribution}
\end{figure*}
%\end{figure}
 
%\subsection{Feature Extraction and Selection}
%The time-varying nature of the frequency responses of active devices necessitates data mining of the VNA's captured traces. In order to correctly extract the relevant features from the traces, we identify the frequency ranges that contain the majority of the information from the traces and are of importance. Note that not every frequency point on the traces is significant, as some of them may be vulnerable to noise or are caused by the device's parasitic capacitance. This is especially true for the traces' high-frequency data points, which are dominantly impacted by the resonance frequency of device components. 

%Figure~\ref{fig:flow} illustrates the categorization process in a simplified form. In order to make use of any machine learning model, an appropriate dataset is required that is composed of the fingerprints of the activities carried out by the firmware. The steps involved in this process are outlined in the subsequent subsections.
%\begin{figure}[htbp]
%\centerline{
%\includegraphics[width= 0.5\textwidth]{Figures/features2.png}}
%    \caption{Proposed feature selection algorithm.}
%    \label{fig:flow}
%\end{figure}

\subsection{Signal Processing}
\label{sec:data_procrssing} \noindent 
%As discussed in Section~\ref{sec:selected_regions}, 
To determine relevant frequency points, we compute the Pearson correlation coefficient between the impedance signal at each frequency and the program instruction set. This allows us to select a subset of the 10,001 total frequency points where the impedance responses are highly correlated with changes in instruction execution. Frequency points with lower correlation to instruction changes are rejected. 

The impedance value at a particular frequency is denoted as $X$ and the instruction set as $I$. For $n$ sample pairs \{($X_1, I_1$),...,($X_n, I_n$)\}, the Pearson correlation coefficient $\rho_{XI}$ is calculated as:
\begin{align}
    \rho_{XI} = \frac{ n \sum X_{i} I_{i} - \sum X_{i} \sum I_{i} }{\sqrt{n\sum X_{i}^{2} - (\sum X_{i})^{2}}   \sqrt{n\sum I_{i}^{2} - (\sum I_{i})^{2}}    } \label{eq:pearson}
\end{align}

We select the largest set of frequencies where the impedance responses are fairly orthogonal, allowing us to further reduce the number of points. Using Equation~\ref{eq:pearson}, we calculate the Pearson correlation coefficients between $X_i$ (impedance at one frequency) and $Y_i$ (impedance at another frequency).
We choose the maximum number of frequencies where the intra-group impedance responses are less than 85\%. In other words, the impedance variations at rejected frequencies can be explained by the impedance at selected frequencies. 
Since impedance at the chosen frequencies can explain the rejected points, we termed these the \mbox{\textit{Dominant frequency points}}. 

Using this two-step frequency selection, we locate approximately 3600 and 2700 frequency points for the FPGA and ATmega328P, respectively, that adequately characterize the software instructions. Notably, these represent just 36\% and 27\% of the total observable frequency points.
We then collect additional impedance signals during various program executions at the specified frequencies to construct the test dataset. By selecting dominant orthogonal frequencies, we drastically reduce the data required while retaining the information needed to reverse-engineer software instructions.

%This section focuses on selecting the appropriate frequencies to investigate for signatures generated from various firmware activities. We desire to find the optimal frequency points in order to reduce the computation time and increase the number of relevant sample points. The procedure is depicted graphically in Figure~\ref{fig:freq_sel}.

\subsection{Feature Extraction}
\label{sec:features_extraction} \noindent 
%As indicated in Figure~\ref{fig:flow}, 
We use the training data to identify features associated with each instruction's execution. We perform principal component analysis (PCA)~\cite{jackson2005user} on the signals to extract these features and then input them to the machine learning (ML) classifiers. 

To reduce training time and increase accuracy, we select a subset of principal components that account for 95\% of the variance in the training data. Using these key components as features improves performance of the classifiers. Figure~\ref{fig:indv_features_fpga} and Figure~\ref{fig:indv_features_atmega} show the distribution of the two most significant principal components for the FPGA and microcontroller instructions, respectively. The axes correspond to the data projected onto the top two components.

% PCA extracts features from the training data and using a subset of the most significant principal components reduces dimensionality and improves classifier accuracy. 

\subsection{Performance Metrics}
\label{sec:metrics} \noindent 
To evaluate classifier performance, we observe five key metrics: recall rate (true positive rate), specificity (true negative rate), precision, accuracy, and F1-score~\cite{marsland2011machine}. All classes are weighted equally when calculating the macro average of these metrics. Let $TP_{i}$, $TN_{i}$, $FP_{i}$, and $FN_{i}$ represent the true positives, true negatives, false positives, and false negatives predicted by a given classifier for class $i$. The metrics are defined as: 
%% The recall rate is defined as the proportion of events that are correctly classified as true relative to the total number of true events, while the specificity rate is defined as the proportion of events that are correctly identified as negative relative to the entire number of actual negatives. 
%False positive (FP) rate is the proportion of actual negatives events that are incorrectly recognized as positive. It is also referred to as a \mbox{``Type I error"}. False negative (FN) rate is the proportion of events incorrectly labeled as negatives relative to the total number of true positives. This is also called a \mbox{``Type II error"}. 
%% Precision rate relates to the predicted accuracy. It is the proportion of events accurately classified as positive relative to the total number of events identified as positive. 
%% Accuracy is the ratio that describes the number of events accurately classified as either truly positive or truly negative out of the total number of events. Note that accuracy is not a particularly relevant metric when test samples are not identical (an unbalanced dataset). 
%% F1-score measures the efficacy of the classification by giving equal weight to recall and precision. This is the harmonic mean between precision and recall. All of these test scores should be close to 1 in an ideal scenario. 
\begin{align}
    Recall = \frac{1}{i} \sum_{i} \left(\frac{TP_{i}}{TP_{i} + FN_{i}}\right) \label{eq:TP} \\
    Specificity = \frac{1}{i} \sum_{i} \left(\frac{TN_{i}}{TN_{i} + FP_{i}} \right)\label{eq:TN} \\
    Precision = \frac{1}{i} \sum_{i} \left(\frac{TP_{i}}{TP_{i} + FP_{i}}\right) \label{eq:precision} \\
    Accuracy = \frac{1}{i} \sum_{i} \left( \frac{TP_{i} + TN_{i}}{TP_{i} + TN_{i} + FP_{i} + FN_{i}}\right) \label{eq:accuracy} 
    \end{align}
    \begin{align}
    F1-score = \frac{2 \times Precision \times Recall}{Precision+Recall} \label{eq:f1} 
\end{align} 
We use Equation~\eqref{eq:TP}-\eqref{eq:f1} to calculate the performance metrics and compare the performance of the classifiers.

\subsection{Detection and Classification} 
\noindent 
We use a number of ML models, including support vector machine (SVM), adaptive boosting (AdaBoost), k-nearest neighbors (kNN), linear discriminant, gaussian naive bayes~\cite{marsland2011machine}. These models are trained to detect and classify program instructions. We utilize 10-fold cross-validation during training to improve stability, avoid overfitting and select the top performer based on its 10-fold cross-validation score.

\begin{table*}[htbp]
\renewcommand{\arraystretch}{1}
\caption{Classification scores for detecting program instruction.}
\label{tab:inv_result_table}
\centering
\begin{tabular}{|c|c||c|c|c|c|c|c|}
\hline
\bfseries Device & Classifier~\cite{marsland2011machine} & Validation Score & F1-Score & Recall & Specificity & Precision & Accuracy \\ %macroAVG
\hline\hline
\multirow{4}{*}{FPGA} 
& SVM (Kernel: Linear)       & 92.8\%           & 92.6\%    & 92.7\% & 99.3\%      & 92.7\%     & 92.6\% \\ \cline{2-8}
& AdaBoost                   & 91.8\%           & 92.0\%    & 92.0\% & 99.3\%      & 92.1\%     & 92.0\% \\ \cline{2-8}
& Linear Discriminant        & 86.6\%           & 87.2\%    & 87.3\% & 98.8\%      & 87.3\%     & 87.3\% \\ \hline \hline
%& KNN                        & 84.5\%           & 85.0\%    & 85.3\% & 98.7\%      & 85.4\%     & 85.1\% \\ \hline \hline
\multirow{4}{*}{ATmega328P} 
& SVM (Kernel: Quadratic)    & 95.0\%           & 96.1\%    & 96.1\% & 99.6\%      & 96.1\%    & 96.1\%     \\ \cline{2-8}
& Bagged trees               & 92.5\%           & 91.6\%    & 91.7\% & 99.2\%      & 91.6\%    & 91.6\%     \\ \cline{2-8}
& Linear Discriminant        & 88.9\%           & 92.2\%    & 92.0\% & 99.2\%      & 92.8\%    & 92.2\%     \\ \hline
%& KNN                        & 84.2\%           & 86.0\%    & 85.8\% & 98.6\%      & 87.4\%    & 85.8\%     \\ \hline
\end{tabular}
\end{table*}

\begin{figure*}[htbp]
\centering
\begin{subfigure}{0.49\textwidth}
    \raggedright
    \includegraphics[width=0.97\textwidth]{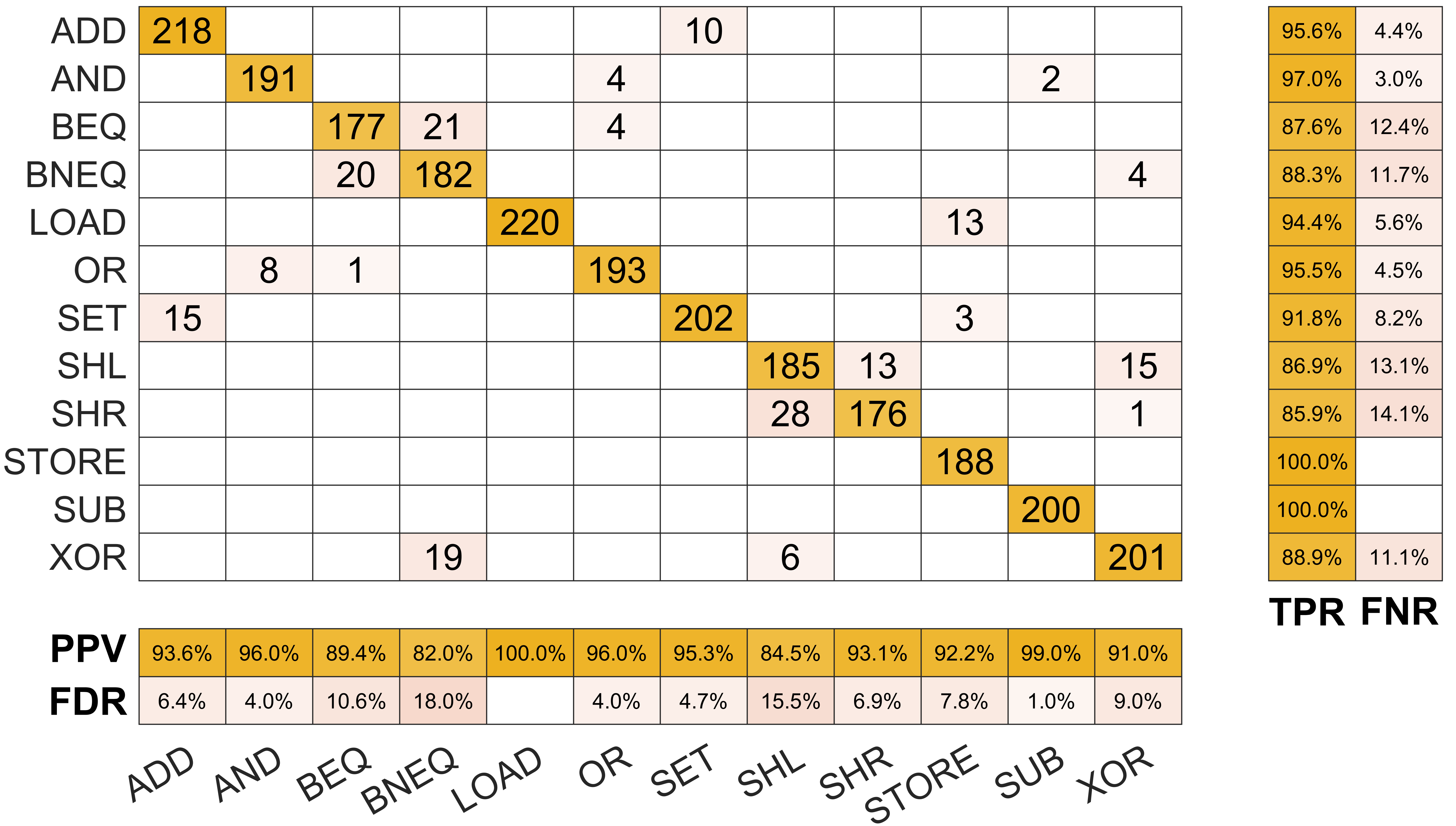}
    \caption{}
    \label{fig:indv_confusion_fpga}
    %\vspace{20pt}
\end{subfigure}
\begin{subfigure}{0.49\textwidth}
    \raggedleft
    \includegraphics[width=0.97\textwidth]{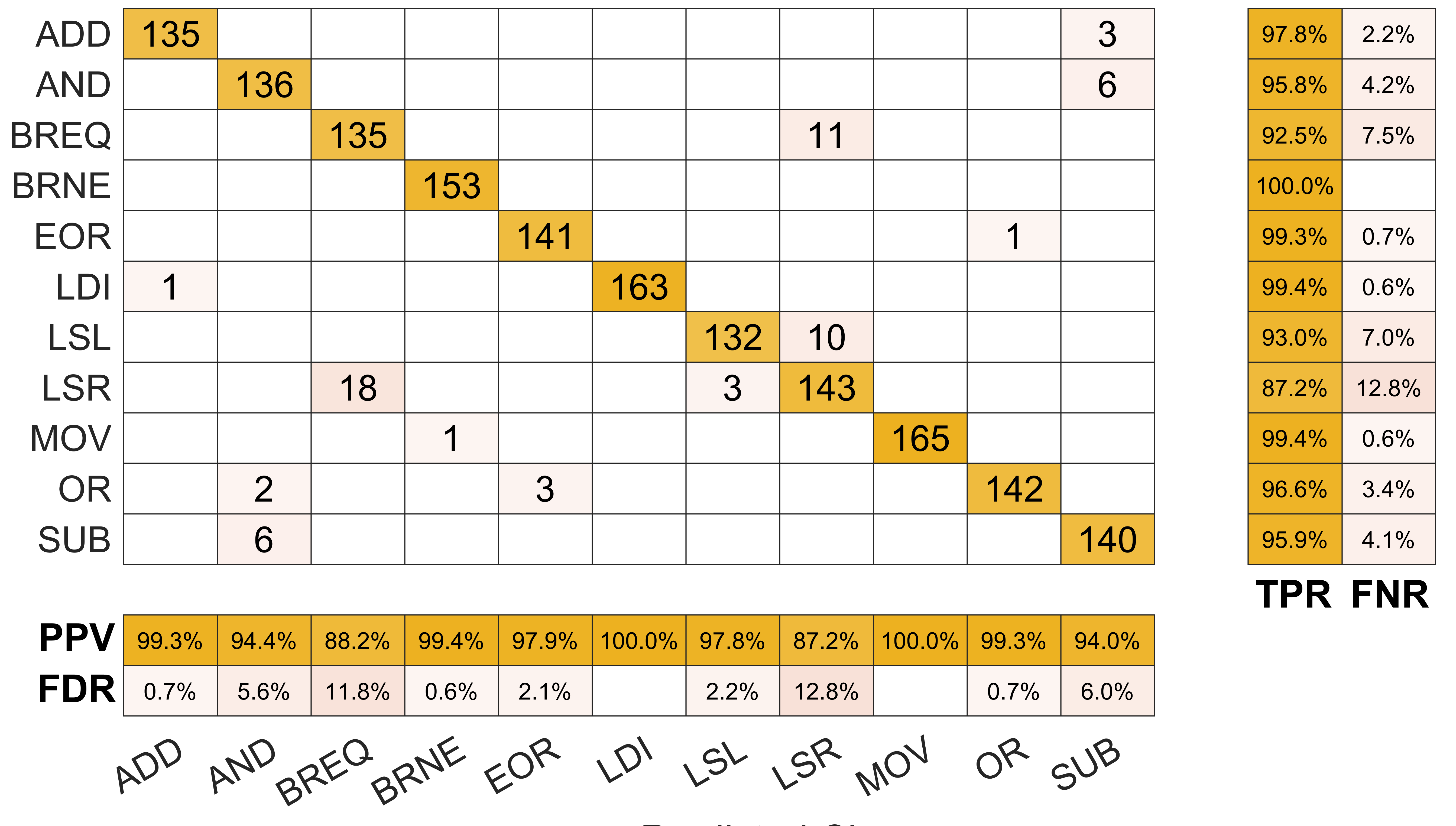}
    \caption{}
    \label{fig:indv_confusion_atmega}
\end{subfigure}
\caption{Confusion matrix of reverse-engineered program instructions: (a) Artix 7 FPGA, (b) ATmega328P microcontroller.}
\label{fig:confusion_inv}
\end{figure*}

After extracting instruction-specific features from the signals as outlined in previous sections (Section~\ref{sec:data_procrssing}, Section~\ref{sec:features_extraction}), the models are evaluated on a 30\% held-out test set. The SVM performs exceptionally well and achieves top classification performance, with F1 score, precision, recall, specificity, and accuracy all over 92\% on the test data. This demonstrates the viability of our proposed approach using impedance side channel for reverse engineering program instructions. The findings of the classification of individual program instructions are presented in Table~\ref{tab:inv_result_table}.  
The confusion matrix in Figure~\ref{fig:confusion_inv} provides further insight into the performance of the top classifier, SVM, 
along with the positive predictive value (PPV), false discovery rate (FDR), true positive rate (TPR), and false negative rate (FNR). PPV represents the probability that a predicted result is actually correct, whereas FDR is the complement of PPV. A high PPV indicates the reliability of the trained model, and the false positive rate is the complement of specificity. Among the other classifiers, SVM demonstrates the ability to accurately differentiate between the various instructions with minimal confusion and a top 10-fold cross-validation score. 

In summary, by leveraging ML algorithms and extracted signal features, our method of using impedance as a side channel can reliably identify and reverse-engineer program instructions executed on a chip. The exceptional classification results from the SVM highlight the promise of this technique. 

% With an F1 score, precision, recall, specificity, and accuracy of more than 92\%, the SVM classifier performs exceptionally well in classifying the program instructions, demonstrating that our method to reverse-engineer the program instructions is viable. 

% check
%The Table~\ref{tab:inv_result_table} displays the five performance metrics of Section~\ref{sec:metrics} for classifying the specific program instructions. The SVM kernel, quadratic SVM (QSVM), cubic SVM, and quadratic discriminant are the models with the highest performance, according to our analysis.
%We find that classifying the type of instruction is more accurate than classifying the instruction itself. With an F1 score, precision, recall, specificity, and accuracy of more than 90\%, the classifiers perform exceptionally well in classifying the types of program instructions, demonstrating that our method to detect the type of program instructions is viable.

\section{Related Works}
\label{sec:related_work}
\noindent 
\textit{\textbf{Side Channel Analysis in Hardware Defense:}}
Side channels in hardware have been leveraged both for extracting sensitive information and securing systems. For example, the authors in~\cite{maji2021leaky} use timing-based information leakage and simple power analysis to analyze vulnerabilities in embedded neural networks. 
In~\cite{he2017hardware}, an EM side channel-based hardware Trojan detection is proposed by spectrum modeling and analyzing. 
Impedance can be employed as a tool for detecting malicious system as well as other physical side channel attacks, e.g., power side channel attacks. 
Authors in~\cite{zhu2023pdnpulse}, detect board-level modifications by measuring impedance at various locations. Another approach in~\cite{fujimoto2018demonstration} detects Trojans by monitoring impedance changes in integrated circuit wires. Authors in \cite{munny2021power} present a detection system that monitors battery impedance to identify malicious probing attempts in power side channel attacks. By detecting impedance changes, their approach provides a new way to thwart such attacks through impedance-based sensing. 
Thus, prior works have demonstrated side channels like timing, power, EM emissions, and impedance can be exploited for analyzing hardware vulnerabilities, detecting Trojans, and identifying malicious modifications. Our research similarly uses impedance as a side channel, but with the novel goal of reverse engineering software instructions.

\noindent
\textit{\textbf{Side Channel Analysis in Software Defense:}}
Recent work has leveraged physical side channels produced by embedded micro-controllers to infer internal behavior and operations. For example, authors in~\cite{basu2019theoretical} present an analytical approach to assess the security of hardware performance counters (HPCs) for malware detection. Their framework analyzes the probability of malware detection given a program, set of HPCs, and sampling rate. 
%By modeling control flows with Hidden Markov Models on an 8-bit AVR, authors in~\cite{liu2016code} measure voltage drops on the power pin to track code execution and detect abnormalities. 
In~\cite{clark2013wattsupdoc}, the authors present a method utilizing power side channel and ML on statistical trace features to detect run-time malware in medical devices. 
% Authors in~\cite{msgna2014precise} utilize instruction-level power side channel leakage profiles for detecting program instructions executed on an embedded processor. 
Analyzing EM emissions is another approach explored in prior art~\cite{han2017watch} detected malicious programmable logic controller (PLC) code by training a sequential neural network model on pre-processed frequency data to encode control flow transitions. 
Other works include % using localized EM probes for dynamic code recognition~\cite{strobel2015scandalee}, 
detecting behavioral changes in industrial control systems~\cite{van2018side}, and identifying malicious PLC code~\cite{boggs2018utilizing}. 
Beyond EM and power analysis, recent work has examined backscattering side channels modulated by impedance variations~\cite{nguyen2019creating}. It should be noted that the backscattering side channel in the paper exploits variations in the signal reflected back from the device. These variations in the reflected signal are caused by changes in the impedance. \\
\noindent
\textit{\textbf{Side channel Analysis in Code Reverse Engineering:}} 
The authors in \cite{eisenbarth2010building} implement an instruction classifier for the PIC16F687 microcontroller by performing a template attack using the power side channel. With statistical models like hidden Markov models (HMMs), they achieve 70.1\% accuracy over 35 test instructions. 
By modeling control flows with HMMs on an 8-bit AVR, authors in~\cite{liu2016code} measure voltage drops on the power pin to track code execution and detect abnormalities. Authors in~\cite{msgna2014precise} utilize instruction-level power side channel leakage profiles for detecting program instructions executed on an embedded processor. Other works include using localized EM probes for dynamic code recognition~\cite{strobel2015scandalee}. 

The extensive prior art demonstrates side channels can reveal rich details about a device's internal operations. Our work similarly leverages impedance as a side channel, but with the unique goal of reverse engineering software instructions on FPGA and microcontroller platforms. By contrast, most existing techniques focus on malware detection, behavior identification, and code profiling rather than reverse engineering the actual opcodes. The ability to reliably reconstruct sequences of software instructions will open up new possibilities for observation, verification, and reverse engineering of embedded and cyber-physical systems.

\section{Conclusion}
\label{sec:conclusion}
\noindent 
In this paper, we investigated the potential of using impedance side channel to identify program instructions. We used hardware platforms including an Artix 7 FPGA and ATmega328P microcontroller, each with a unique instruction set. By analyzing the impedance signals, we generated fingerprints and employed machine learning classifiers. We found specific instructions could be identified with 92.6\% accuracy on the FPGA and 96.1\% on the microcontroller. 
This research pioneers the use of impedance as an active side channel for hardware security. It shows device impedance can reveal executed instructions, opening new possibilities for code recognition, code flow analysis, code reverse engineering and embedded systems security. Analyzing the relationship between impedance and software represents an important advance. It promises to revolutionize techniques for strengthening resilience against cyberthreats in an increasingly connected world. In conclusion, transforming impedance from a passive, overlooked property into an active, potent side channel marks a critical step forward for the field.

% Our results demonstrate that the impedance side channel can disassemble program instructions and be used as a monitoring system to protect an embedded system from executing malicious program instructions, verifying program integrity, detecting malware, and even detecting counterfeit systems. %However, further research is required to evaluate their robustness in real-time systems. 

\section*{Acknowledgement}
\noindent
% We would like to thank the reviewers for providing their valuable feedback.
This work is supported partly by the National Science Foundation Award \# 2114200 and the National Security Agency Award \# H98230-22-1-0327.

\normalem
\bibliographystyle{IEEEtran}
\bibliography{IEEEabrv,references}
\end{document}